\documentclass[12pt, a4paper]{article}
\pdfoutput=1

\usepackage{amsmath}
\usepackage{amsfonts}
\usepackage{amssymb}
\usepackage{graphicx, rotating}
\usepackage{epsfig}
\usepackage{array}
\usepackage{latexsym}
\usepackage{graphicx}
\usepackage{xcolor}
\usepackage{amsmath,bm,amssymb}
\usepackage{slashed}
\usepackage{hyperref}
\usepackage[normalem]{ulem}
\usepackage[normalem]{ulem}
\usepackage[utf8]{inputenc}
\usepackage[export]{adjustbox}
\usepackage{multirow, verbatim}
\usepackage{ulem}

\usepackage{overpic}

\newcommand{\be}{\begin{equation}}
\newcommand{\ee}{\end{equation}}
\newcommand{\bea}{\begin{eqnarray}}
\newcommand{\eea}{\end{eqnarray}}
\newcommand{\nn}{\nonumber}
\newcommand{\total}{\mathrm{d}}

\begin{document}

%FRONTPAGE2%%%%%%
%\pagenumbering{gobble}
\begin{titlepage}

%\vspace{-1.6in}
\begin{flushright}
\small
DESY-21-109
\end{flushright}
\vspace{.3in}

\begin{center}
{\Large\bf Finding sound shells in LISA mock data \\ using likelihood sampling} \\
\bigskip\color{black}
\vspace{1cm}{
  {\large
Felix~Giese,
Thomas Konstandin,
Jorinde~van~de~Vis
}}

{\small
Deutsches Elektronen-Synchrotron DESY, Notkestr. 85, 22607 Hamburg, Germany
}
\bigskip

\begin{abstract}
We study to what extent LISA can observe features of gravitational 
wave spectra originating from cosmological first-order phase transitions.
We focus on spectra which are of the form of double-broken power laws. 
These spectra are predicted by hydrodynamic simulations and also analytical models such as 
the sound shell model. We argue that the ratio of the two break frequencies is 
an interesting observable since it can be related to the wall velocity 
while overall amplitude and frequency range are often degenerate for the numerous 
characteristics of the phase transition. Our analysis uses mock data obtained from the power spectra predicted by the simplified simulations and the sound shell model and analyzes the detection prospects using $\chi^2$-minimization and likelihood sampling. We point out that the prospects of observing 
two break frequencies from the electroweak phase transition is hindered by a shift of the spectrum to smaller frequencies for strong phase 
transitions. Finally, we also highlight some 
differences between signals from the sound shell model compared to simulations.
\end{abstract}

\end{center}

\end{titlepage}
%\pagenumbering{arabic}

\tableofcontents

%%%%%%%%%%%%%%%%%%%%%%%%%%%%%%%%%%%%%%%%%%%%%%%%%%%%%%%%%%

\newpage
%%%%%%%%%%%%%%%%%%%%%%%%%%%%%%%%%%%%%%%%%%%%%%%%%%%%%%%%%%
\section{Introduction}
\label{sec:Introduction}
%%%%%%%%%%%%%%%%%%%%%%%%%%%%%%%%%%%%%%%%%%%%%%%%%%%%%%%%%%

The gravitational wave telescope LISA \cite{2017arXiv170200786A} will probe stochastic gravitational wave (GW) backgrounds with unprecendented precision in the near future. A possibly observable signal is formed by gravitational waves originating from a cosmological first-order phase transition (PT). This possibility received a lot of attention after the discovery of the Higgs,
because the electroweak phase transition -- triggered by the Higgs field~-- is first-order in many models that feature an extended scalar sector or strong couplings.

While current collider experiments at the LHC test properties of the Higgs field, such as its mass, its couplings to Standard Model (SM) particles 
and its vacuum expectation value (VEV), the information from GW experiments are complementary. 
For example, the characteristics of the phase transition are rather sensitive to dark sectors coupling to the Higgs~\cite{Schwaller:2015tja} which cannot be easily tested at collider experiments. 

The information of the phase transition is encoded in the shape of the GW spectrum that
features a peak in the mHz regime in case the PT took place around electroweak temperature scales~\cite{Grojean:2006bp}. The most accurate predictions of the gravitational wave signal today have been obtained from lattice simulations of the bubble interacting with the plasma \cite{Hindmarsh:2013xza, Hindmarsh:2015qta, Hindmarsh:2017gnf}. 
Although the spectrum can have more than one characteristic feature, it is often described by a broken power law with a single peak frequency, see for example 
the recent review~\cite{Caprini:2019egz}.

The most important characteristics in this expression are the temperature of the PT (denoted by $T^*$), the wall velocity
during the first-order PT (denoted by $v_w$), the inverse duration of the PT (denoted by $\beta$) and the amount of energy released into the 
plasma by the phase transition (denoted by $\alpha$, see below for more comments on the quantity). So in total there 
are four underlying parameters 
\be
\label{eq:fourP}
\alpha, \, \beta, \, T^*, \, v_w.
\ee

This leads to the unfortunate conundrum that one would typically observe only two features (peak position and amplitude, whilst the two slopes of the spectrum are expected to be universal) of the GW spectrum,
which results in degeneracies in the underlying four characteristics. In particular, the peak frequency 
depends on $T^*$ and $v_w/\beta$ while the peak amplitude depends on $\alpha$, $v_w$ and $\beta$.
Hence, additional assumptions have to be made for example by fixing the PT temperature $T^*$ and the PT duration $\beta^{-1}$
to specific values, see e.g.~\cite{Gowling:2021gcy}.

Fortunately, it is reasonable to expect that the GW spectrum has a more complicated structure than the formula from Ref.~\cite{Caprini:2019egz}. The reason is that there are several relevant length scales in the process. One break in the spectrum is related to the average separation between bubbles towards the end of the PT when most GWs are produced, denoted by $R_*$, which is related to the phase transition duration and bubble wall velocity via
\be
	R_* = \frac{(8\pi)^{1/3} v_w}{\beta}.
\ee
In addition, the PT features a second length scale that might be imprinted as a second break in the spectrum: the thickness of the 
sound shells $L_{\rm ssh}$, which is the size of the region around the bubble wall where the fluid is in motion. One finds
\be
H^{-1} \gg R_* \gtrsim L_{\rm ssh} \, .
\ee
When $L_{\rm ssh}$ differs significantly from $R_*$, the GW spectrum will be a double-broken power law
rather than a broken power law with a single peak. Interestingly, the ratio of these two break frequencies 
allows to measure the ratio $R_*/L_{\rm ssh}$ which in turn depends mostly on the wall velocity $v_w$ and 
to a lesser extent on the strength of the phase transition $\alpha$~\cite{Gowling:2021gcy}. 
This makes it an excellent target to learn more about 
the characteristics of the PT even though a small degeneracy in the other parameters (and possibly an ambiguity in the wall velocity) still persists. In this work, we will simulate the LISA detection of a GW spectrum with two characteristic length scales (see e.g. Eq.~(\ref{eq:hybsim})) and study how much information can be extracted from such a spectrum.

Unfortunately, the amount of data from full-fledged hydrodynamic simulations is still somewhat 
too limited to study the behavior of the sound shell thickness $L_{\rm ssh}$ in detail.
In the present work, we will therefore model the source of GWs in two alternative ways. We will employ the sound shell 
model as given in~\cite{Hindmarsh:2019phv} and also fits to the simplified simulations provided in~\cite{Jinno:2020eqg}.
We will see that there are significant differences in the prediction for $L_{\rm ssh}$ in its dependence
on the four parameters given in Eq.~(\ref{eq:fourP}). 

Recently, a study of the observational prospects of LISA was presented in~\cite{Gowling:2021gcy}. One of the main results of this
study is that indeed the double broken power law shape is essential and allows for a reliable measurement of the wall velocity. 
In comparison, the focus of the present study is more on the question what are the prospects of observing the double broken power law shape 
in the first place. Compared to the study in~\cite{Gowling:2021gcy}, our setup is somewhat more general: we consider the sound shell model as well as results from simplified simulations to model the expected GW spectrum. Since both models differ considerably in their predictions, we fit to a six parameter model that leaves the slope of the source spectrum towards the IR and UV unspecified
and is not optimized to fit the predictions of either of the two source models we use.
Instead of treating the temperature and phase transition temperature as external parameters, we use 
concrete models from the literature to relate these two parameters to the strength of the phase transition.
As we will see, this leads to the unfortunate observation that for strong phase transitions, one of the breaks in the 
spectrum is often shifted out of the LISA sensitivity band~\cite{Huber:2007vva}.  
To estimate the uncertainties, we use a Markov chain Monte Carlo instead of the Fisher matrix, which in combination with 
the more general fitting model leads to some sizeable deviations. In particular, we find some non-elliptical 
probability contours.   
Some additional comments on how our work relates to~\cite{Gowling:2021gcy} are given in the discussion section.

Our modeling of the sources are discussed in Section \ref{sec:Sources} and \ref{subsec:AIC}.
In Section \ref{sec:LISA} we discuss how we generate mock data for our LISA analysis and in Section \ref{sec:Analysis}
we set up the analysis using $\chi^2$ as well as likelihood approaches. 
In this section, we also present our results and we conclude in Section \ref{sec:Conclusion}.
The Appendices provide additional information on the implementation of the sound shell model and details on the Markov chain Monte Carlo parameter reconstruction.

%%%%%%%%%%%%%%%%%%%%%%%%%%%%%%%%%%%%%%%%%%%%%%%%%%%%%%%%%%
\section{Sources of the GW spectrum}
\label{sec:Sources}
%%%%%%%%%%%%%%%%%%%%%%%%%%%%%%%%%%%%%%%%%%%%%%%%%%%%%%%%%%

In this section we desribe how we model the GW spectrum as a function of the four parameters
given in Eq.~(\ref{eq:fourP}). As discussed in the introduction, our main concern is the 
double-broken structure in the GW spectrum that arises when the scales $R_*$ and $L_{\rm ssh}$ do
not coincide. 

This feature has not yet been studied systematically in the literature. In particular, in hydrodynamic
simulations, the observation of this feature is technically challenging since besides the 
bubble separation $R_*$ and the shell thickness $L_{\rm ssh}$ also the wall thickness has to be resolved 
in the simulation. This limits the range where $R_*/L_{\rm ssh}$ is large enough to be observable and at the same 
time does not interfere with the other length scales (such as the box size and the grid resolution). 

\subsection{Simplified simulations}

Simplified simulations~\cite{Jinno:2020eqg} of the phase transition have closed this gap to a certain extent. Additional assumptions 
(like e.g.~linearity of the sound waves) have to be accepted for these simulations. Still, the results seem to be in 
quantitative agreement with full hydrodynamic simulations and allow to study the impact of the sound shell thickness.

In~\cite{Jinno:2020eqg}, both breaks in the spectrum have been observed and correlated with the naive thickness of the sound 
shells that can be read of from the fluid profiles of spherical bubbles before collision. It was argued that the shell thickness resulting \emph{after} collision is related to the thickness in spherically symmetric simulations via
\be
L_{\rm ssh} \simeq 
( \xi_{\rm front} - \xi_{\rm back} )/ \beta = \xi_{\rm shell}/ \beta \, ,\label{eq:lss}
\ee
where $\xi_{\rm front/back}$ denote the beginning and end of the non-zero fluid profile for the different expansion modes (detonations/deflagrations/hybrids). 
In~\cite{Jinno:2020eqg}, a comparison between this heuristic and the simulation result found qualitative agreement. Nevertheless, this was 
only tested for relatively weak phase transitions ($\alpha < 0.05$).

For the source at production time we use the formulas presented in~\cite{Jinno:2020eqg}
\bea
\Omega^*_{\rm gw, sim} = 5 \, \xi_{\rm shell} \, (\kappa\alpha)^2 \, S(f^*) \frac{4 H^* \tau}{3\pi^2} \frac{H^*}{\beta} \, .\label{eq:sourcesim}
\eea
We put $^*$s on quantities to denote that they are evaluated at the time the gravitational wave spectrum is produced. $f^*$ thus denotes the frequency at production time and $S(f^*)$ is the spectral shape
\be
S(f^*) = N \frac{(f^*/f_1)^3}{1 + (f^*/f_1)^3 + (f^*/f_1)^3 (f^*/f_2)^3} \label{eq:hybsim}\, ,
\ee
with the break positions $f_1 \simeq 1/R_*$ and $f_2 \simeq \beta/(\xi_{\rm front} - \xi_{\rm back})$. The spectrum is normalized such that
$\int df^*/f^* \, S(f^*) = 1$. 

The energy budget $\kappa\alpha$ is determined by solving the relativistic hydrodynamic equations for a single spherical bubble interacting with the plasma~\cite{LandauLifshitz, Steinhardt:1981ct, KurkiSuonio:1995pp, Espinosa:2010hh}. The fit from~\cite{Espinosa:2010hh} can be used to obtain the energy budget as a function of the wall velocity and the phase transition strength $\alpha$ 
\be
	\alpha = \frac{1}{3} \frac{\Delta \theta (T_n)}{w_+(T_n)}\, ,\qquad \theta = e -3 p\, ,
\ee
with $p$ and $e$ the pressure and energy density respectively and $\Delta$ denoting the difference between the broken and symmetric phase. \cite{Giese:2020rtr, Giese:2020znk} generalized the results of~\cite{Espinosa:2010hh} for a general equation of state, but in this work we will assume for simplicity that the equation of state is radiation-like and $c_s^2 = 1/3$. The hydrodynamic equations are also used to determine the mean squared fluid velocities $\bar U_f^2$
\be
	\bar U_f^2 \simeq \frac34 \kappa \alpha\, ,\label{eq:uf}
\ee
where we used the adiabatic index $\Gamma = 4/3$ for a relativistic plasma and assume small $\alpha$.

The factor $\tau$ in Eq.~(\ref{eq:sourcesim}) determines the duration of the GW source that is either cut off by the Hubble 
expansion or the decay of the sound waves into turbulence\cite{Ellis:2020awk}. The decay time into turbulence is 
hereby given as
\be
\tau_{\rm nl} = R_* / \sqrt{\bar U_f^2} \, {\rm max}(1,c_s/v_w)
\ee
and is shorter than the inverse Hubble rate, $\tau \simeq \tau_{\rm nl}$ for all our parameter choices.

\subsection{Sound shell model}
A second possibility to predict a quantitative GW spectrum including the second 
length scale $L_{\rm ssh}$ is to use the sound shell model (SSM)~\cite{Hindmarsh:2016lnk,Hindmarsh:2019phv}. In this model, the velocity 
power spectrum is determined using the whydrodynamic solutions of spherical, isolated bubbles. 
Assuming that the correlators of the fluid velocity are Gaussian, the two-point function of the 
energy-momentum tensor and GW production rate can be deduced. The sound shell gravitational wave spectrum is given in Eq.~(\ref{eq:soundshellsource}) and further details are discussed in Appendix~\ref{sec:Soundshell model}.

The outcome of the sound shell model can be significantly different than what has been seen in simulations. 
Overall, the source seems to be smaller compared to the simulation results.
Besides, the two length scales as observed in the GW spectrum are
well separated for deflagration and hybrid expansions but then almost merge for slightly larger 
wall velocity when the expansion modes transitions to detonations. This suggests that the second break is set by the width of the shock for deflagrations and hybrids, while for detonations the width of the rarefaction wave is the relevant scale.

\subsection{Redshift and models}
The gravitational wave spectra from the simulations are defined in terms of $f^*$, the frequency at the time of production, and the sound shell spectra are defined in terms of the dimensionless quantity $q=kR_*$. To convert the sound shell spectrum into a function of wave number $k$, we divide by $R_*$, 
\be
	\frac{q}{R_*} = \frac{q}{(8\pi)^{1/3}v_w} \frac{\beta}{H^*} H^*\, ,
\ee
and we obtain the corresponding frequency at production as
\be
	f^* = 1.12 \times 10^9 {\rm Hz} \, \frac{1}{v_w} \frac{\beta}{H^*} \left(\frac{T^*}{100 { \rm GeV}} \right)^2 \left(\frac{g_*}{100} \right)^{1/2} q\, .
\ee
To obtain the frequency as observed by LISA $f_0$, we have to account for the redshift
\bea
	f_0 &=& 7.97 \times 10^{-16} \left(\frac{100 { \rm GeV}}{T^*} \right) \left(\frac{100}{g_*} \right)^{1/3} f^*  
\label{eq:redshiftf},
\eea
where $g_*$ is the number of relativistic degrees of freedom at the phase transition temperature. 

The overall amplitude also is redshifted and we obtain
\be
	h^2 \Omega_{\rm gw}(f_0) =1.64 \times 10^{-5} \left(\frac{100}{g_*} \right)^{1/3} \Omega^*_{\rm gw}\left(f^* \right)\, \label{eq:redshiftA},
\ee

Once the model for the GW spectrum (sound shell model or simulation fit) is selected, the observed 
GW spectrum still depends on the the four parameters given in Eq.~(\ref{eq:fourP}).
In order to reduce the complexity of the problem, several assumptions can be made. 
As explained before, for most parts we would like to vary the wall velocity $v_w$ and study 
the ratio of the two break positions in the spectrum resulting from the separation of the scales $R_*$ and $L_{\rm ssh}$.

One reasonable choice would be to fix the phase transition temperature $T^*$ (which will be of electroweak scale $\sim$ 100 GeV) and
the duration of the phase transition $\beta$ (that for models typically is in the range $\beta/H^* \sim 100 - 1000$). 

However, it is known that these parameters actually are strongly correlated with the stength of the 
parameter $\alpha$. Stronger phase transition tend to reduce the PT temperature $T^*$ and also the inverse duration $\beta/H^*$.
This leads to the unfortunate observation that stronger phase transitions eventually escape detection since the GW spectrum shifts
towards the IR and out of the sensitivity band of LISA~\cite{Huber:2007vva}.

In the following we would like to capture this effect and hence assume not fixed values but fixed relations via the phenomenological relations
\be
T^*/{\rm GeV} = 48.4 - 33.8 \alpha + \frac{0.259}{(0.0677 + \alpha)^2} \, \label{eq:Tfroma},
\ee
and
\be
\beta/H^* = 26.3 + 74.8\alpha + \frac{8.85}{(0.0592 + \alpha)^2} \, \label{eq:bfroma}.
\ee
These relations are approximately true in the two Higgs doublet model for ratios of Higgs VEVs, $\tan\beta_v\simeq 10$, and have been obtained by fitting the datapoints provided on {\tt PTPlot} \cite{Caprini:2019egz}. Quite similar relations are 
found for singlet extensions of the Standard Model and the extension with effective higher-dimensional operators~\cite{Caprini:2019egz, Schmitz:2020rag}.
Most of our analysis will rely on these relations but we will eventually increase the temperature by a factor 10 
to mimick electroweak symmetry breaking from strong dynamics~\cite{Randall:2006py}.

\begin{figure}
\centering
\includegraphics[width = 0.48\textwidth]{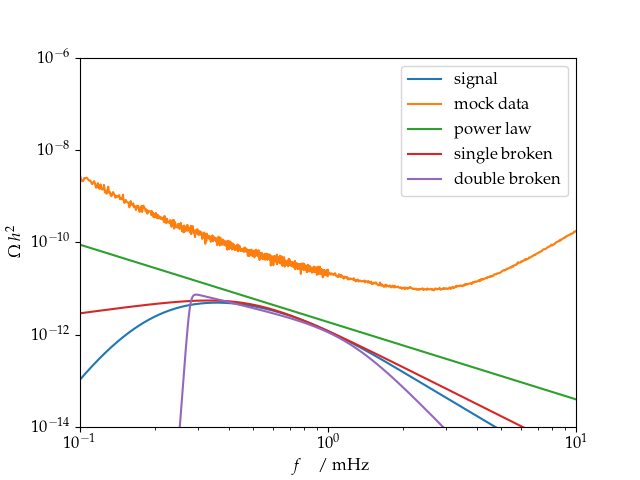}
\includegraphics[width = 0.48\textwidth]{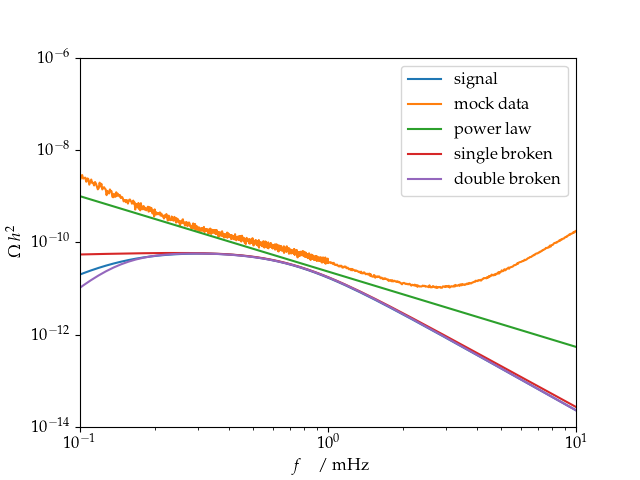}
\caption{Example data and fits with $\alpha = 0.25$, $v_w=0.4$. The left plot shows a signal from the sound shell model and the right plot shows a signal from simplified simulations. For both cases, the single broken power law is preferred (according to the criterion discussed in Section~\ref{subs:chisq}).}
\label{fig:exs}
\end{figure}

%%%%%%%%%%%%%%%%%%%%%%%%%%%%%%%%%%%%%%%%%%%%%%%%%%%%%%%%%%
\section{LISA mock data generation}
\label{sec:LISA}
To generate the LISA mock data, we follow the approach of \cite{Caprini:2019pxz}. We briefly summarize the approach here. 

We assume that the LISA mission will last 4 years, in which data are collected 75\% of the time. Due to the need of regular operational breaks, the data will be collected in 94 chunks, with a corresponding frequency resolution of $\Delta f = 10^{-6} \rm Hz$. 

To good approximation,  the noise signal is characterized by
\be
	\Omega_{\rm noise}(f) = \frac{4 \pi^2}{3 H_0^2} f^3 \frac{\frac{10}{3}\left(1+ 0.6 \left(\frac{2 \pi f L}{c} \right)^2 \right)\left(P_{\rm oms} (f,P) + \left( 3 + \cos{\frac{4\pi f L}{c}}\right)P_{\rm acc}(f,A) \right)}{(2 \pi fL/c)^2}\, ,
\ee
where $L$ is the arm length and $c$ the speed of light. $P_{\rm oms}$ and $P_{\rm acc}$ are the power spectral densities from the optical metrology system and the mass acceleration, respectively:
\bea
	P_{\rm oms} (f,P) &=& P^2 \frac{\rm pm^2}{\rm Hz} \left(1 + \left(\frac{2 \rm \, mHz}{f} \right)^4 \right) \left(\frac{2\pi f}{c} \right)^2,  \\
	P_{\rm acc} (f,A)&=& A^2 \frac{\rm fm^2}{\rm s^4 Hz} \left(1 + \left(\frac{0.4 \rm \, mHz}{f} \right)^2 \right) \nn \\
&& \times \left(1 + \left(\frac{f}{8 \rm \, mHz} \right)^4 \right) \left(\frac{1}{2\pi f} \right)^4 \left(\frac{2\pi f}{c} \right)^2,
\eea
where $P\sim 15$ and $A \sim 3$ are noise parameters. 

We consider frequencies between $3 \times 10^{-5} \rm \, Hz$ and $0.5 \rm \, Hz$ with a spacing of $\Delta f = 10^{-6} \rm \, Hz$. For each frequency $f_i$, we generate $N_c = 94$ data points for the different chunks (labeled by $j$), as a sum of the signal and noise contributions
\be
	D_{i,j} = S_{i,j} + N_{i,j}\, ,
\ee
where the $S_{i,j}$ and $N_{i,j}$ are obtained from the power spectra:
\begin{align}
	&S_{i,j} = \left \rvert \frac{G_{i1}\left(0,\sqrt{h^2 \Omega_{\rm gw}(f_i)}\right) + iG_{i2}\left(0,\sqrt{h^2 \Omega_{\rm gw}(f_i)} \right)}{\sqrt 2} \right\rvert^2 , \\
	& N_{i,j} = \left \rvert \frac{G_{i3}\left(0,\sqrt{h^2 \Omega_{\rm noise}(f_i)}\right) + iG_{i4}\left(0,\sqrt{h^2 \Omega_{\rm noise}(f_i)} \right)}{\sqrt 2} \right\rvert^2.
\end{align}
The numbers $G_{ik}(0,\sigma)$ are randomly drawn from a Gaussian distribution with zero mean and standard deviation set by the square root of the power spectra. For each frequency, we obtain the mean $\bar D_i$ and standard deviation $\sigma_i$ from the 94 data points.

To reduce the computational load, we coarse grain the simulated data for frequencies larger than $10^{-3} \, {\rm \, Hz}$. The data with frequencies between $10^{-3}\, {\rm \, Hz}$ and $0.5 \,{\rm \, Hz}$ are grouped into 1000 bins. The signal $D$ and frequency $f$ of the bins is obtained by a weighted average of all the $\bar D_i$ and $f_i$ in the bin, with weights $1/\sigma_i^2$. The overall standard deviation of the bin is given by
\be
	\sigma = \left(\sum_i \frac{1}{\sigma_i^2}\right)^{-1/2}\, .
\ee

The coarse graining procedure of Ref.~\cite{Caprini:2019pxz} was modified in Ref.~\cite{Flauger:2020qyi}, to remove a bias in the reconstructed parameters coming from non-Gaussianities.
Since the $D_{i,j}$ follow a $\chi^2$ distribution, the sample means $\bar D_i$ and sample variances $\sigma_i^2$ are correlated. This leads to an additional bias and ultimately a mismeasurement of the noise parameters. This issue will be treated later.

%%%%%%%%%%%%%%%%%%%%%%%%%%%%%%%%%%%%%%%%%%%%%%%%%%%%%%%%%%
\section{Analysis}
\label{sec:Analysis}
%%%%%%%%%%%%%%%%%%%%%%%%%%%%%%%%%%%%%%%%%%%%%%%%%%%%%%%%%%
In this section, we study the ability of LISA to characterize the gravitational wave signal. Even though the input gravitational wave signal might be best described by a double-broken power law, (part of) the signal might be unrecoverable due to noise. As a result, a better fit (see the discussion around Eq.~(\ref{eq:AIC}) for our criterion) might be provided by a model with fewer parameters. In this situation, LISA can not extract the maximal amount of information from the gravitational wave signal.

Our analysis works as follows. After generating mock data for a range of phase transition parameters we pretend to be agnostic about the input signal, and attempt to reconstruct the gravitational wave spectrum. The analysis consists of two parts. In the first part we study the ability of LISA to observe a signal and to identify its break(s). In the second part we study more carefully what accuracy can be obtained for the reconstructed parameters.

\subsection{Parameter scan with $\chi^2$}\label{subs:chisq}
In the first step, we reconstruct the gravitational wave spectrum by minimizing the $\chi^2$: 
\be
	\chi^2(\vec\theta_m, \vec\theta_n) = N_c \sum_i \left[\frac{\bar D_i - h^2 \Omega_{\rm gw}(f_i,\vec\theta_m) - h^2 \Omega_{\rm noise}(f_i,\vec \theta_n)}{\sigma_i} \right]^2,\label{eq:chi2}
\ee
as a function of the model parameters $\vec\theta_m$ and noise parameters $\vec\theta_n = {P,A}$. We minimize $\chi^2$ for four different models (with reference frequency $f_0 = 1$ mHz):
\begin{itemize}
	\item{{\bf No gravitational wave spectrum.} $h^2 \Omega_{\rm gw} =0$, i.e. we fit the signal to a spectrum that consists of noise only.}
	\item{{\bf Power law}
		\be
			h^2\Omega_{\rm gw}(f, B, p_1) = B \left(\frac{f}{f_0}\right)^{p_1}\, .
		\ee
	}
	\item{ {\bf Broken power law }
		\be
			h^2\Omega_{\rm gw}(f, B, f_1, p_1,p_2) = B \frac{\left(\frac{f}{f_0}\right)^{p_1}}{1+\left(\frac{f}{f_1}\right)^{p_2}}\, .
		\ee
	}
		\item{ {\bf Double-broken power law }
		\be
			h^2\Omega_{\rm gw}(f, B, f_1,f_2, p_1,p_2,p_3) = B \frac{\left(\frac{f}{f_0}\right)^{p_1}}{\left(1+\left(\frac{f}{f_1}\right)^{p_2}\right)\left(1+\left(\frac{f}{f_2}\right)^{p_3}\right)}\, \label{eq:fitdb}.
		\ee
	}
\end{itemize}
We do not resort to the fit functions of~\cite{Caprini:2019egz,Hindmarsh:2019phv}, in which the shape around the peak and the IR- and UV-slopes are fixed, in order to avoid overfitting. This reflects the reality that in the actual LISA analysis, the true shape of the spectrum will not be known. Besides, we have to accommodate the shape of both source models in our analysis.

In contrast to~\cite{Caprini:2019pxz}, we do not split our data set into smaller bins, but instead we minimize the $\chi^2$ of the entire data set at once.
We determine the overall best fit by comparing the Akaike Information Criterion (AIC)~\cite{1100705} of the four different fits:
\be
	{\rm AIC} = \chi^2_{\rm best fit} + 2 k\, \label{eq:AIC},
\ee
where $k$ is the number of fit parameters. A comparison between the AICs instead of the bare $\chi^2$ prevents overfitting.

\subsection{Comparison of the AIC}\label{subsec:AIC}
\begin{figure}[t]
\centering
\includegraphics[width = \textwidth]{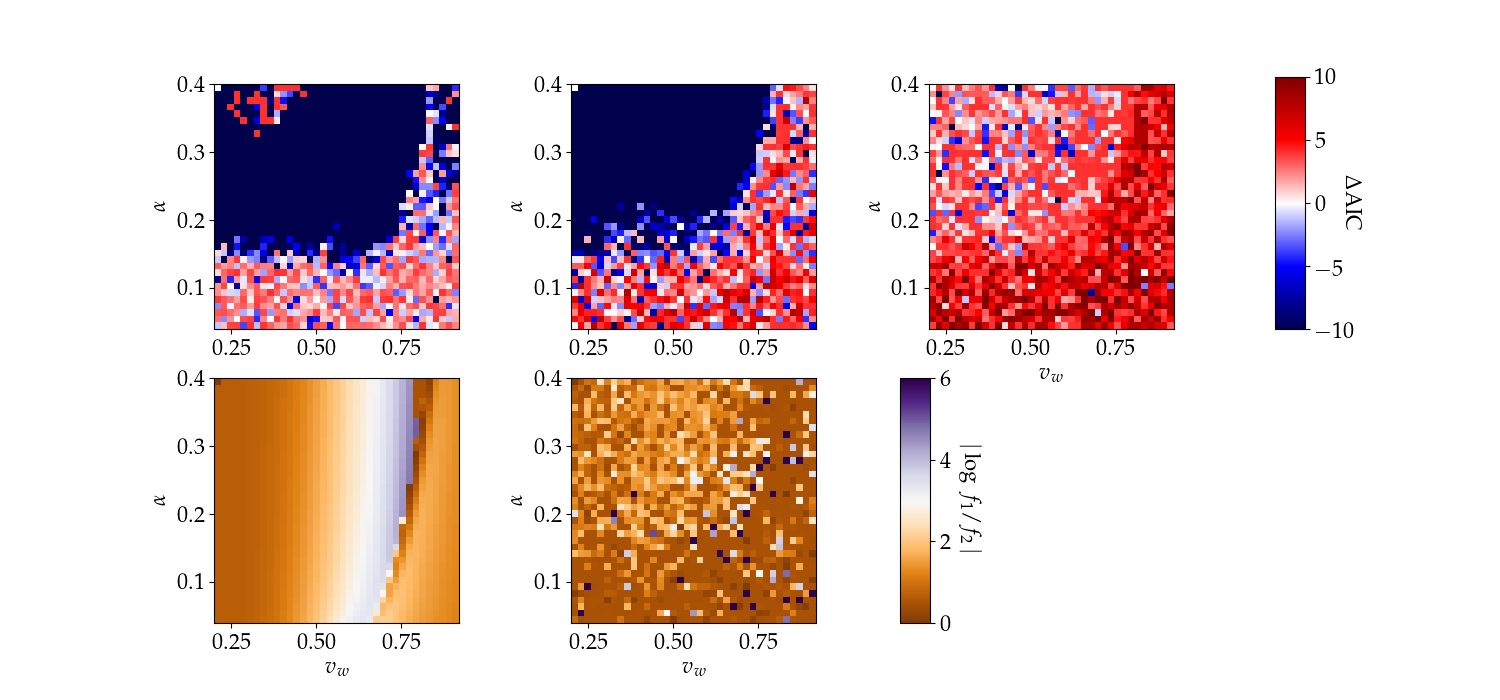}
\caption{Analysis of the fit quality for the sound shell model, with the relation between $\alpha$ and $T^*$ and $\beta/H^*$ set by Eqs.~(\ref{eq:Tfroma},\ref{eq:bfroma}).
Top row: comparison of the AIC of the double broken power law (right), single broken power law (middle) and (power law) with the minimum AIC of fits with fewer parameters for the sound shell model. A positive AIC, signalling a poorer fit, is indicated by red, and a negative AIC, signalling a better fit, is indicated by blue. 
Bottom row: logarithm of the ratio of the break positions $|\log{f_1/f_2}|$, obtained from the input signal (left) and the fit to the double broken power law (right).
}
\label{fig:AICss}
\end{figure}

\begin{figure}[t]
\centering
\includegraphics[width = \textwidth]{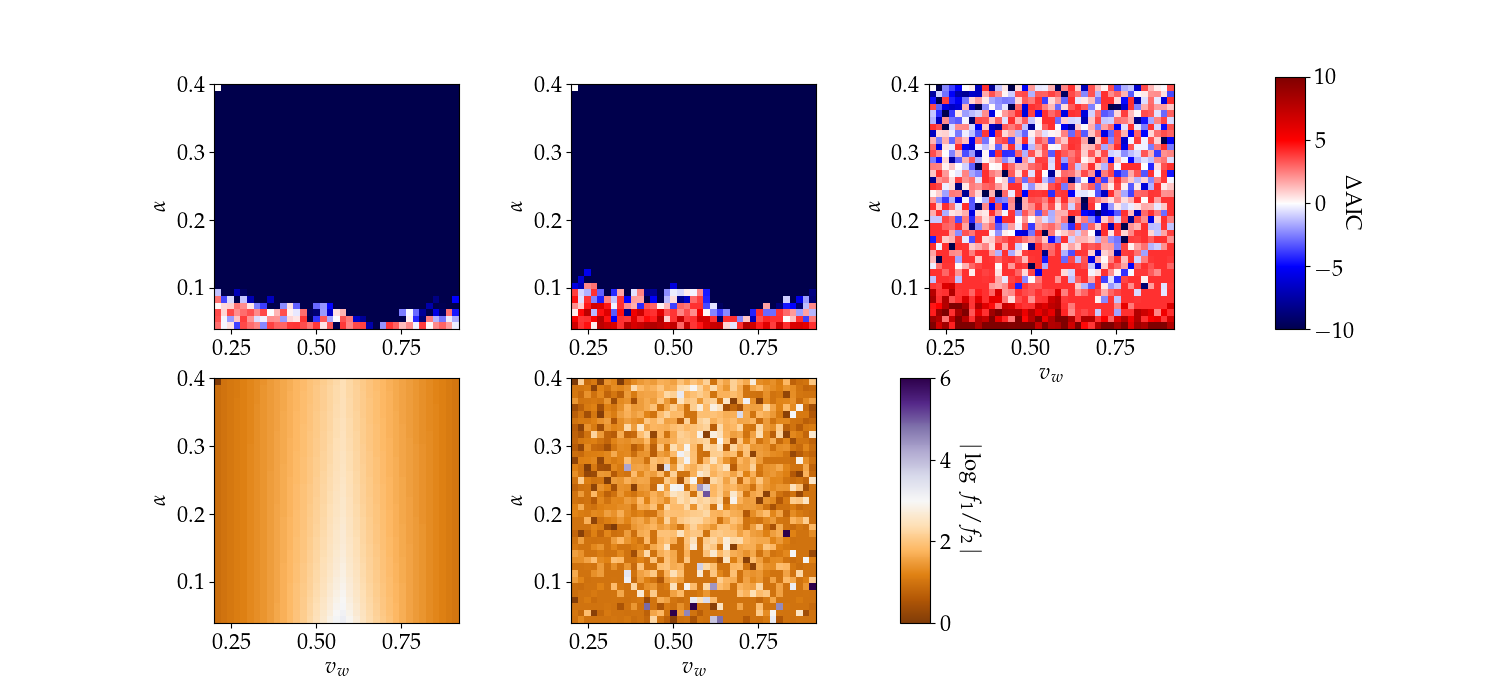}
\caption{The same quantities as in Figure~\ref{fig:AICss}, but for the simulation data, with the relation between $\alpha$ and $T^*$ and $\beta/H^*$ set by Eqs.~(\ref{eq:Tfroma},\ref{eq:bfroma}).}
\label{fig:AICsim}
\end{figure}

\begin{figure}
\centering
\includegraphics[width = \textwidth]{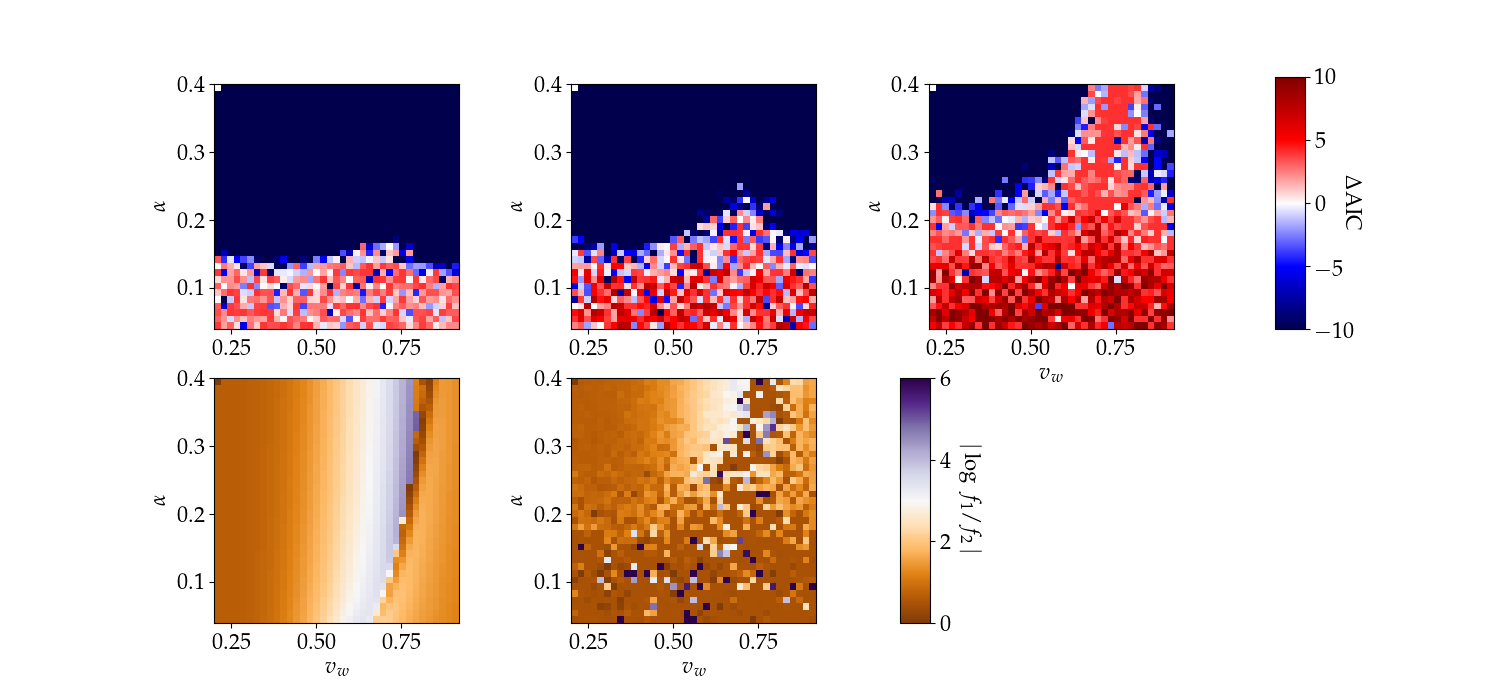}
\caption{The same quantities as in Figure~\ref{fig:AICss} for the sound shell data. The relation between $\alpha$ and $\beta/H^*$ is set by Eq.~(\ref{eq:bfroma}), but the temperature is increased by a factor 10.}
\label{fig:AICssT}
\end{figure}

\begin{figure}
\centering
\includegraphics[width = \textwidth]{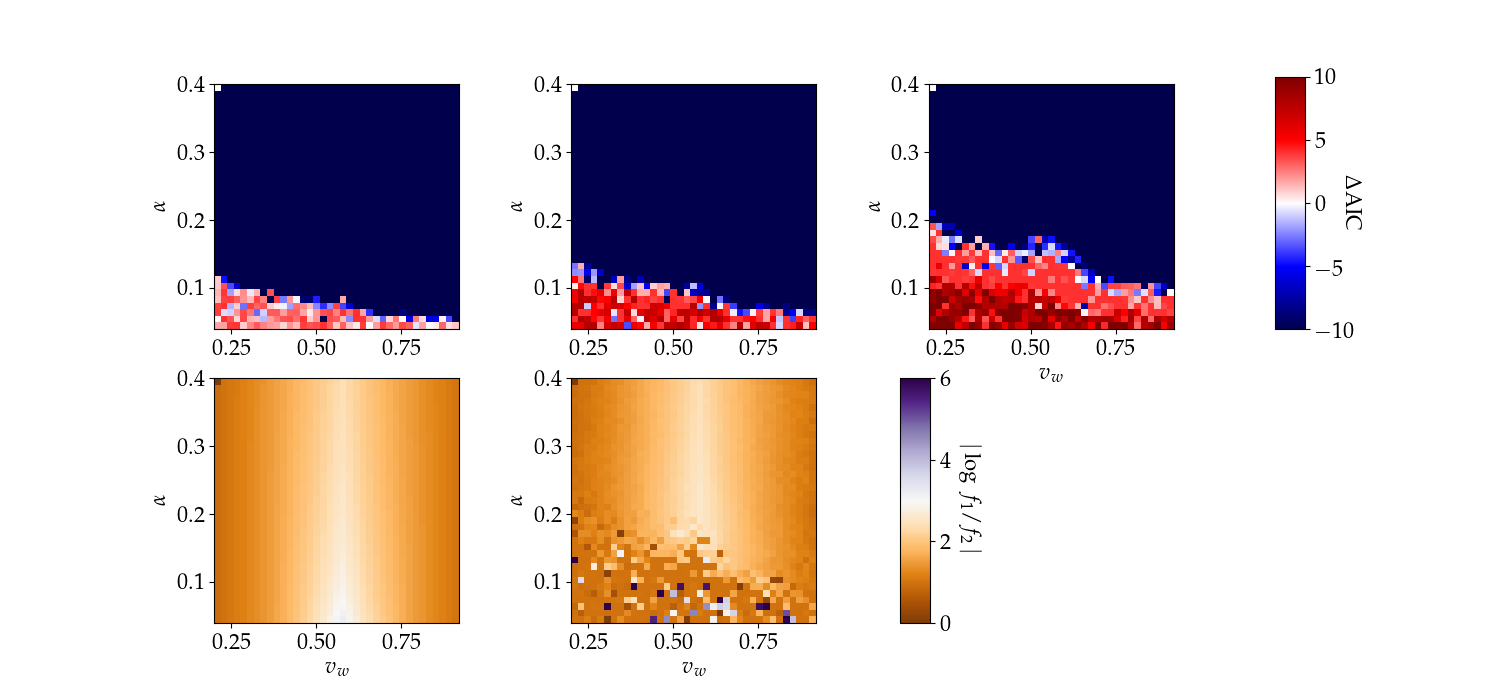}
\caption{The same quantities as in Figure~\ref{fig:AICss}, but for the simulation data. The relation between $\alpha$ and $\beta/H^*$ is set by Eq.~(\ref{eq:bfroma}), but the temperature is increased by a factor 10.}
\label{fig:AICsimT}
\end{figure}

We present our results of the $\chi^2$ comparison in Figures~\ref{fig:AICss} to \ref{fig:AICsimT}. In the top row of each figure we compare the AIC of a given model (right: double broken power law, middle: single broken power law, left: power law), with the smallest AIC of the models with fewer parameters. When this difference is positive (red), we conclude that the model with fewer parameters provides the better fit. 

In the bottom row of Figures~\ref{fig:AICss} to \ref{fig:AICsimT} we show the distance between the two breaks{breaks}. The left plots show $|\log{f_1/f_2}|$, which is obtained by mapping the input signal onto Eq.~({\ref{eq:fitdb}}). This plot can be used as a benchmark for the relation between the ratio of  the break positions and the wall velocity. The right plot shows the reconstruction of the distance between the two breaks from a $\chi^2$ fit of the LISA mock data including the signal and noise.

Figures~\ref{fig:AICss} and \ref{fig:AICsim} demonstrate our results for the sound shell model and the simulations. We vary $\alpha$ and $v_w$ between $0.04$ -- $0.4$ and $0.2$ -- $0.92$ respectively. The values of $T^*$ and $\beta/H^*$ are obtained from Eqs.~(\ref{eq:Tfroma},\ref{eq:bfroma}). The right plot in the top row of Figure~\ref{fig:AICss} demonstrates that the double broken power law fit typically has a larger AIC than fits with fewer parameters for the sound shell model, meaning that the two break positions cannot be reconstructed. The appearance of some blue points and the relatively small positive value of $\Delta {\rm AIC}$ in the upper left corner of the right graph demonstrate that these are really borderline cases; the double break can almost be reconstructed. The middle plot demonstrates that in that region of parameter space, with relatively small bubble wall velocity and large phase transition strength, the single broken power law signal provides the best fit. The explanation for this is that the low-frequency break of the input signal lies at a frequency where LISA's sensitivity is not optimal. The fit therefore only reconstructs the high frequency break. There is only a small region, with large $\alpha$ and $v_w$, for which the power law provides the best fit, and a large range (the region that is mostly red in all three graphs) for which noise only provides the best fit, and no signal is reconstructed at all.

The detection of a signal is significantly more likely for the data predicted by the simulations. There is a small region for strong phase transitions and slow wall velocities where the double broken power law is reconstructed.\footnote{Note that simulations in~\cite{Cutting:2019zws} demonstrate that vorticity is produced in this region of parameter space, which is not captured by the simplified simulations.} 
For phase transitions with $\alpha \gtrsim 0.15$ the single broken power law provides the best fit for all wall velocities. In a thin sliver around $\alpha \sim 0.1$, the power law provides the best fit. Only for $\alpha\lesssim0.1$, the reconstruction finds only noise.

The bottom left plots of Figures~\ref{fig:AICss} and \ref{fig:AICsim} provide further insight in the relation between the break distance and the wall velocity, and also in the difference between the two source models. The relation between the break distance and the wall velocity is already suggested by Eq.~(\ref{eq:lss}), which sets the position of one of the breaks, and which depends on $\beta$ and on $v_w$ via $\xi_{\rm shell}$. As the other break position is set by $R_*$, the $\beta$-dependence drops out of the break ratio. The bottom left plot of Figure~\ref{fig:AICsim} indeed shows a clear relation between the break distance and the wall velocity, implying that the reconstructed break distance could be used to infer the value of the bubble wall velocity. Distinguishing between the two values of the velocity that yield the same break ratio requires input from the other reconstructed signal parameters and/or input from the new physics model.

The break distance becomes largest when the wall velocity approaches the speed of sound, since the sound shell becomes very thin in this regime. The bottom middle graph of Figure~\ref{fig:AICsim} shows that the reconstruction of the break distance is remarkably successful for the simulation data, given the limited success of the double broken power law fit indicated by the top right graph.

The relation between the break distance and the wall velocity for the sound shell model, shown in Figure~\ref{fig:AICss} looks quite different from the relation of the simulation data. The difference is most apparent at the transition from hybrids to detonations. The explanation is that in the SSM the position of the second break is set by the width of the shock, see Figure~\ref{fig:soundshell01}, which suddenly disappears at the Jouguet velocity, leading to an abrupt change in the spectrum. In the simplified simulations, on the other hand, the bubble collisions are simulated, and these wash out the sharp shock front of the hybrids (see e.g.~Figure~14 of~\cite{Jinno:2020eqg}). As a result, the rarefaction wave gives the dominant contribution to the gravitational wave spectrum, and no discontinuity appears at the Jouguet velocity. 

We now keep the relation between $\alpha$ and $\beta/H^*$ the same as in Eq.~(\ref{eq:bfroma}), but increase the temperature by a factor 10 compared to Eq.~(\ref{eq:Tfroma}). This increase in the temperature is not completely ad hoc, since models like composite Higgs and gauged lepton models feature phase transition temperatures around the TeV-scale. As becomes clear from Eqs.~(\ref{eq:redshiftf}, \ref{eq:redshiftA}), the only effect of the rescaling of the temperature is a shift of the spectrum to higher frequency. The results are shown in Figures~\ref{fig:AICssT} (sound shell model) and \ref{fig:AICsimT} (simulations). 

For both cases, there is now a significant region of parameter space in which the double broken power law can be reconstructed. For the sound shell model it is the parameter space at small wall velocity and relatively strong phase transitions, and a small region with large wall velocity and strong phase transition. For the simulations it also includes the region of fast wall velocities and intermediate to strong phase transitions. The bottom rows demonstrate that the reconstruction of the break distance is very successful in the regime where the two breaks are observed. 

\subsection{Markov chain Monte Carlo}\label{sec:MCMC}

We will now employ a more careful reconstruction of the double-broken structure in the sound shell model. As the double-broken structure is mostly observed in the $T$ shifted spectra, we adopt this shift also in this section. We will use the Markov chain Monte Carlo (MCMC) method to estimate the accuracy of our analysis. Besides, we improve the likelihood function to account for non-Gaussianities in our mock data. 

In the Bayesian analysis the posterior probability density of the model and noise parameters $p(\vec{\theta_m},\vec{\theta_n}|\mathrm{data})$ is proportional to the product of the priors on the fitting parameters $\pi(\vec{\theta_m}), \pi(\vec{\theta_n})$ and the likelihood function $\mathcal{L}(\mathrm{data}|\vec\theta_n,\vec\theta_m)$
\be
p(\vec{\theta_m},\vec{\theta_n}|\mathrm{data})\propto P(\theta_m,\theta_n)\equiv {\pi}(\vec\theta_m)\pi(\vec\theta_n)\mathcal{L}(\mathrm{data}|\vec\theta_m,\vec\theta_n)\, .
\ee 

In this language, one can interpret the $\chi^2$-minimization of the previous subsection as the maximum likelihood estimate (MLE) of the unnormalized posterior $P$
\be
\left.\frac{\partial\log P}{\partial \vec\theta}\right\rvert_{\vec\theta=\theta_{\mathrm{MLE}}}=0\, ,
\ee
with the Gaussian likelihood function $\log\mathcal{L}=\log\mathcal{L}_{\chi^2}=-2\chi^2$, and flat priors
\be
\partial_{\vec{\theta_m}}\log\pi(\vec\theta_m)=0\,\qquad\partial_{\vec{\theta_n}}\log\pi(\vec\theta_n)=0\, . 
\ee 

The $\chi^2$-fit only gives correct MLEs for Gaussian distributed data. Since our mock data is not Gaussian- but $\chi^2$-distributed, we introduced a systematic bias of the order of the skewness of the sample, when obtaining the model parameters by $\chi^2$-minimization. The fact that the $\sigma_i$ enter as weights in the coarse graining of the data, is another source of systematic bias. We follow Ref.~\cite{Flauger:2020qyi} to account for this and instead of $\sigma_i^2$ use the following weight in the coarse-graining procedure of a data point in bin $k$   
\be
w_i (f_i) = \frac{(h^2\Omega_{\rm noise}(f_i))^{-1}}{\sum_{j \in {\rm bin }\, k} (h^2\Omega_{\rm noise}(f_j))^{-1}}\, ,\label{eq:cgfancy}
\ee
where the variances are now given by the noise spectra $h^2\Omega_{\rm noise}$. This ensures that the coarse grained data does not suffer from the correlation between $\bar D_i$ and $\sigma_i$.

\begin{table}[t]
	\centering
	\begin{tabular}{|c|c|c|c|c|}
		\hline 
		$v_w$ & $\alpha$ & $\Delta f_\mathrm{signal}$ & $\Delta f_\mathrm{\chi^2}$ & $\Delta f_\mathrm{\mathrm{G+LN}}$  \\ 
		\hline 
		\hline 
		0.3 & 0.4 & 0.7397 & 0.7554& 0.7323  \\ 
		\hline
		0.4 & 0.25 & 0.9948 & 0.9657& 0.9531  \\ 
		\hline 
		0.5 & 0.30 & 1.5795& 1.5042 & 1.5546\\ 
		\hline 
		0.92 & 0.3 & 1.2942 & 1.5825 &1.6492  \\ 
		\hline 
	\end{tabular} 
	\caption{Chosen points for the MCMCs. For the two set-ups, the chains are initialized around the respective maximum likelihood points. The values for the respective break ratio observable $\Delta f\equiv\log f_1/f_2$ at the point of maximum likelihood are given above. The fit to the signal is given for comparison.}
	\label{tab:MCMCbenchmarks}
\end{table}

Further, we write down a new likelihood function that accounts for the non-Gaussianities in the fit $\mathcal{L}_\mathrm{G+LN}$ \cite{Flauger:2020qyi}
\be
\label{eq:Lfull}
\log \mathcal{L}_\mathrm{G+LN}\equiv \frac{1}{3} \log \mathcal{L}_\mathrm{G}+\frac{2}{3}\log \mathcal{L}_\mathrm{LN}\, ,
\ee
where
\be
\log\mathcal{L}_\mathrm{G}=-\frac{N_c}{2}\sum_i n_{f,i}\left(\frac{\bar{D_i}-h^2\Omega_{\mathrm{noise}}(f_i)-h^2\Omega_{\mathrm{GW}}(f_i)}{ h^2\Omega_{\mathrm{noise}}(f_i) + h^2\Omega_{\mathrm{GW}}(f_i) }\right)^2\, ,
\ee
and we included a log-normal contribution to account for the slightly non-Gaussian nature of the data
\be
\log\mathcal{L}_\mathrm{LN}=-\frac{N_c}{2}\sum_i n_{f,i} \log^2 \frac{h^2\Omega_{\mathrm{noise}}(f_i) + h^2\Omega_{\mathrm{GW}}(f_i)}{\bar D_i}\, .
\ee
Note that it is the weighting that differs from the $\chi^2$ of Eq.~(\ref{eq:chi2}), as the variance is now estimated from the theoretical uncertainty instead of the data itself, consistent to the updated coarsegraining method. The sum $i$ is over coarse grained data, and $n_{f,i}$ denotes the number of frequencies in each bin. 

To determine the accuracy of the $\chi^2$ analysis from the previous step of the analysis we set up two MLEs. The first estimate closely resembles the previously discussed $\chi^2$-fit, using the MLE \footnote{Note that the subscript $\chi^2$ indicates that the likelihood is directly proportional to $\chi^2$, not that it is given by a $\chi^2$-distribution.}
\be
\log \mathcal{L}_{\chi^2}=-2\chi^2\, .
\ee
The second one takes non-Gaussianities, better uncertainty estimates, as well as prior information on the noise parameters into account. For this 
we use the likelihood function $\mathcal{L}_\mathrm{G+LN}$ given in Eq.~(\ref{eq:Lfull}).

For the signal parameters $\vec{\theta_m}$, we adopt flat priors in both cases,
\be
\log \pi(\vec{\theta_m})=0\, .
\ee 
For the noise parameters $A$,$P$ we adopt either flat ($\delta_{\chi^2}=1$) or Gaussian ($\delta_{\chi^2}=0$) priors, according to Refs.~\cite{Caprini:2019pxz,Flauger:2020qyi}, 
\be
\log \pi(\vec{\theta_n})=(1-\delta_{\chi^2})\sum_{\theta_n}\frac{(\theta_n-\theta_{n,\mathrm{center}})^2}{{(\theta_{n,\mathrm{center}}/5)^2}}\, ,\,\mathrm{with}\qquad \vec{\theta}_{n,\mathrm{center}}=(A,P)=(3,15)\, .
\ee
Note that this prior dependence is actually very mild given the data-sample is largely noise-dominated and has no impact on the analysis.

We now choose a couple of selected benchmark points, which according to the $\chi^2$ analysis have a reconstructable double-broken structure (see Figure~\ref{fig:AICssT}). The benchmark points are listed in Table~\ref{tab:MCMCbenchmarks}.

\begin{figure}[t]
	\centering
	\includegraphics[width=0.24\linewidth]{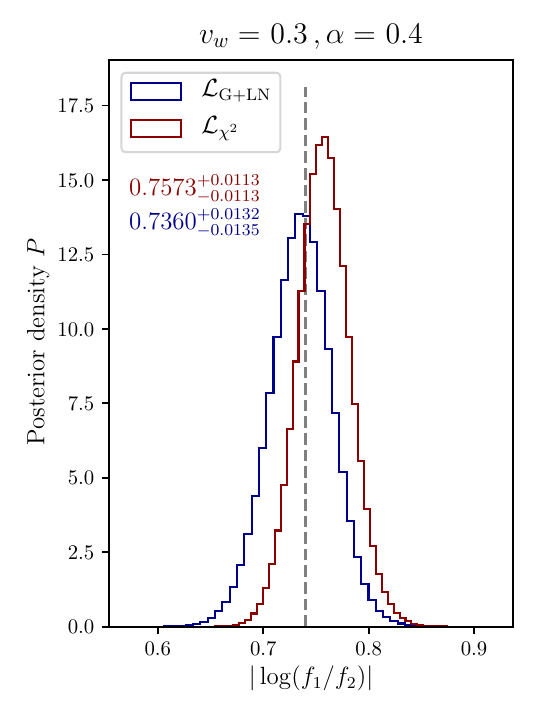}
	\includegraphics[width=0.24\linewidth]{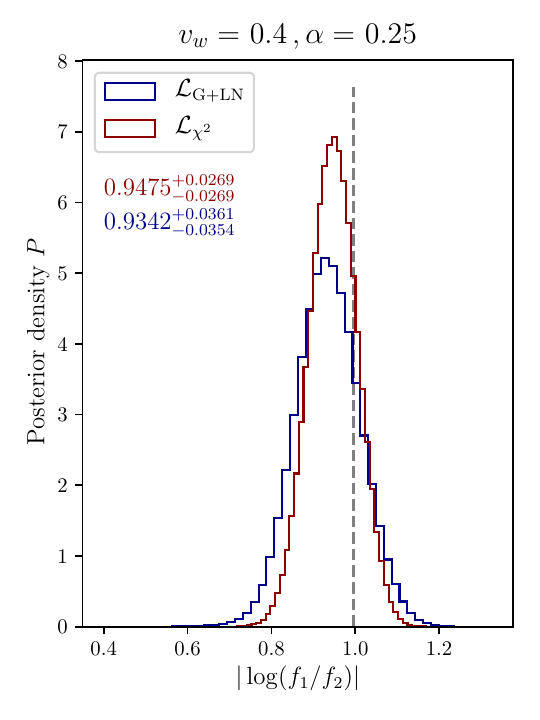}
	\includegraphics[width=0.24\linewidth]{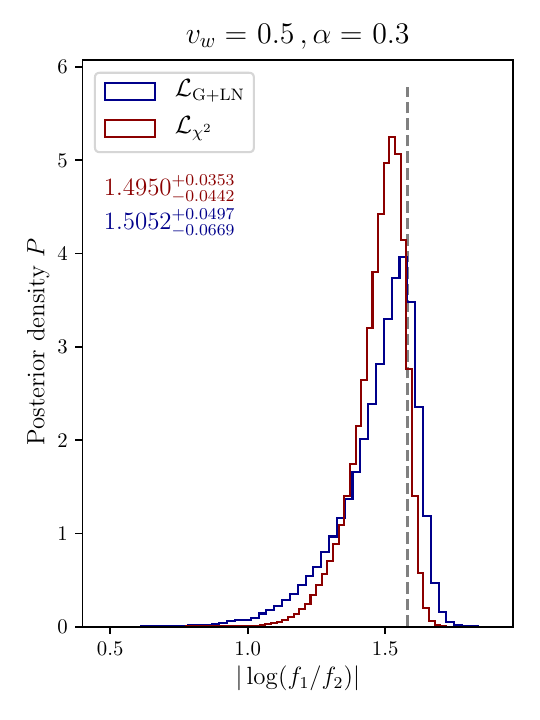}
	\includegraphics[width=0.24\linewidth]{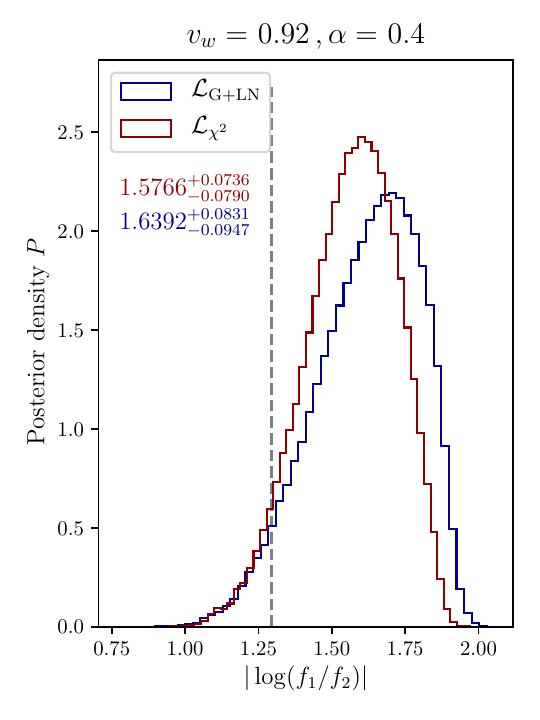}
	\caption{Posterior probability density for four selected parameter points $(v_w, \alpha)$, see Table~\ref{tab:MCMCbenchmarks}. Depending on the point the uncertainty in the reconstruction of the break ratio observable $|\log f_1/f_2|$ is of order $\mathcal{O}(0.1-1)$. 
The dashed lines shows the best-fit value from the input signal.}
	\label{fig:posteriors}
\end{figure}

We sample the two likelihood functions around the MLE using the publicly available Markov chain Monte Carlo (MCMC) code \emph{emcee}~\cite{ForemanMackey:2012ig}. The chain length $N_{\text{MCMC}}$ is chosen such that the autocorrelation $\tau_\text{ac}$ length of all chains fulfills $\sigma_\mathrm{\mathrm{MCMC}}^2\propto\tau_\text{ac}/N_\text{MCMC} \ll 1/50$. Figure~\ref{fig:posteriors} shows the resulting posterior distribution of the break ratio observable $\log(f_1/f_2)$. Further details of the signal reconstruction are shown in Appendix~\ref{sec:appMCMC}.

Figure~\ref{fig:posteriors} and the figures in Appendix~\ref{sec:appMCMC} demonstrate two interesting findings. First, non-Gaussianities seem to play an important role in the analysis. This becomes most obvious in the reconstruction of the noise parameters in Figures~\ref{fig:triangle30alph40000} to \ref{fig:triangle92alph40000}. The noise parameters are systematically underestimated by the $\chi^2$ analysis. Even though the overall qualitative picture stays the same, the other parameter estimates are also impacted. This means that the analysis using the non-Gaussian likelihood function can do better than a simple $\chi^2$ analysis. Moreover, the reconstructions in Figures~\ref{fig:triangle30alph40000} to \ref{fig:triangle92alph40000} reveal several non-elliptical error contours. These features can not be correctly reproduced in a Fischer matrix analysis.

Second, in case the two breaks in the spectrum are resolved according to the AIC, the observable $|\log f_1/f_2|$ can be measured with about $10\%$ accuracy.
Of course, the precision will deteriorate when the border of the detectability of two breaks is approached. Still, the observable $|\log f_1/f_2|$ can  provide interesting and reliable information about the properties of the phase transition when a second break in the spectrum is discovered.

\section{Discussion\label{sec:Conclusion}}

The main target of this work is a dedicated study of the observable 
\be
| \log f_1/f_2 | \, ,
\ee
where $f_1$ and $f_2$ denote two breaks in a GW spectrum as observed by LISA. The motivation
consider this  observable is that it encodes the ratio of two physical scales in case 
the spectrum results from a first-order cosmological phase transition
and mostly depends on the wall velocity during the PT. 

The first scale, $R_*$, corresponds to the mean separation of bubbles towards the end of the 
phase transition, when most of the gravitational radiation is produced. The second scale, $L_{\rm ssh}$,
corresponds to the typical thickness of the sounds shells that are produced during the phase transition.

In order to make connection to actual phase transition characteristics, we used two approaches 
to predict the GW spectrum. The first one is the so-called sound shell model~\cite{Hindmarsh:2016lnk, Hindmarsh:2019phv} and the second relies 
on fits to results from simplifed simulations~\cite{Jinno:2020eqg}.

We observe that the observable behaves quite differently in these two approaches. In the sound shell model, 
the observable is largest for the hybrid expansion mode where a very thin shock preceeds the wall. 
This happens when the wall velocity is slightly below the Jouguet velocity. Once the wall velocity 
surpasses the Jouguet velocity, the observable sharply drops in the sound shell model.
In the simulations, this feature is washed out by the collisions of the sounds shells and the subsequent evolution of the sound
shells. Hence the observable is typically largest when the wall velocity is close to the speed of sound, where the shock front is strongest.

Overall, we expect that the simplified simulations represent the true spectrum better than the sounds shell model, since the simulations account for the evolution of the plasma after bubble collision and do not assume a Gaussian fluid velocity spectrum.
Nevertheless, the simplified simulations have not been tested (and are probably not applicable since linearity of the sound waves is assumed) for strong phase transitions, $\alpha>0.05$. 
In case a double-broken power law will indeed be observed by LISA, a dedicated hydrodynamic simulation with increased precision will be indispensable. Ultimately, we estimate that the above observable can be measured with a relative error below $10\%$ in case
the two breaks are clearly identified in the data. We do point out that the prospect of measuring the break ratio is largest for phase transitions at a relatively high temperature $T^* \sim 1 \, {\rm TeV}$, since for smaller temperature, the low-frequency break 
often lies outside of the LISA sensitivity band. The size of the parameter space for which the two breaks can be observed will increase when GW experiments become even more sensitive.

At this point, we would like to compare our results to the recent study~\cite{Gowling:2021gcy}.
As mentioned before, we focus more on the prospects of observing the double broken power law structure rather than
the general observational prospects. Hence, we use two different approaches to model the source and we also 
use a more general fitting function in the data analysis (leaving the two exponents in the deep IR and UV unspecified).
Another difference is that we use the Markov chain Monte Carlo method to determine the uncertainties in the parameters, while the analysis of~\cite{Gowling:2021gcy} mostly used the Fisher matrix. In contrast to~\cite{Gowling:2021gcy} we find some non-Gaussian 
features in our parameter estimates. This difference is probably explained by the fact that 
we go beyond the Gaussian approximation in the construction of the log-likelihood and use a fit model with more parameters.    
The final main difference is that we relate the phase transition temperature $T^*$ and the inverse duration of the phase transition
$\beta$ to the strength parameter rather than treating them as an external parameters. As discussed before, this leads to a sizeable 
shift towards lower frequencies for strong phase transitions. 

There are also several minor differences that do not have a
large impact on the qualitative results: we assume that the lifetime of the sound waves is determined by decay into turbulence (see Section~\ref{sec:Sources}), neglect the potential suppression in GW sources from deflagrations~\cite{Cutting:2019zws}, neglect foregrounds, and model noise following~\cite{Caprini:2019pxz}. Overall, our results seem to agree quite well whenever a one-to-one comparison is possible.

There are several ways to improve on the above analysis. First, following \cite{Caprini:2019pxz}, we only simulated the data taking of one Time Delay Interferometry channel. An improved analysis would also simulate the other two channels and then switch to the so-called AET basis, in which the signal and noise covariance matrices are diagonal \cite{Prince:2002hp}. This updated analysis corresponds to a $\sqrt 2$ improvement in the signal-to-noise ratio.

Second, we have not included any stochastic gravitational wave foreground signals (see~\cite{Gowling:2021gcy} for a discussion on reconstruction of the sound shell model signal in the presence of foregrounds). One possible source of confusion noise is caused by unresolved extragalactic compact binaries~\cite{Regimbau:2011rp}. The signal strength is expected to lie significantly below the LISA sensitivity curve and thus our analysis should not be significantly affected. The foreground coming from white dwarf binaries \emph{within} our galaxy \cite{1987ApJ...323..129E, Bender_1997}, on the other hand, can be a significant noise source in our frequency range. The proximity of the binary sources possibly allows a distinction between the binary foreground and the GW signal from the phase transition due to the annual modulation of the former \cite{Adams:2013qma}. 

\section*{Acknowledgements}
We thank Mauro Pieroni for help with the data analysis and Chloe Gowling and Mark Hindmarsh for useful comments on the manuscript. This work is supported by the Deutsche Forschungsgemeinschaft under Germany's Excellence Strategy - EXC 2121 Quantum Universe - 390833306.

\newpage
%%%%%%%%%%%%%%%%%%%%%%%%%%%%%%%%%%%%%%%%%%%%%%%%%%%%%%%%%%
\appendix
%\section*{Appendix}
\label{sec:app}
%%%%%%%%%%%%%%%%%%%%%%%%%%%%%%%%%%%%%%%%%%%%%%%%%%%%%%%%%%

%%%%%%%%%%%%%%%%%%%%%%%%%%%%%%%%%%%%%%%%%%%%%%%%%%%%%%%%%%
\section{The Soundshell model}
\label{sec:Soundshell model}
\begin{figure}[t!]
	\centering
	\includegraphics[trim= 0 0 170 0,width=0.99\linewidth,]{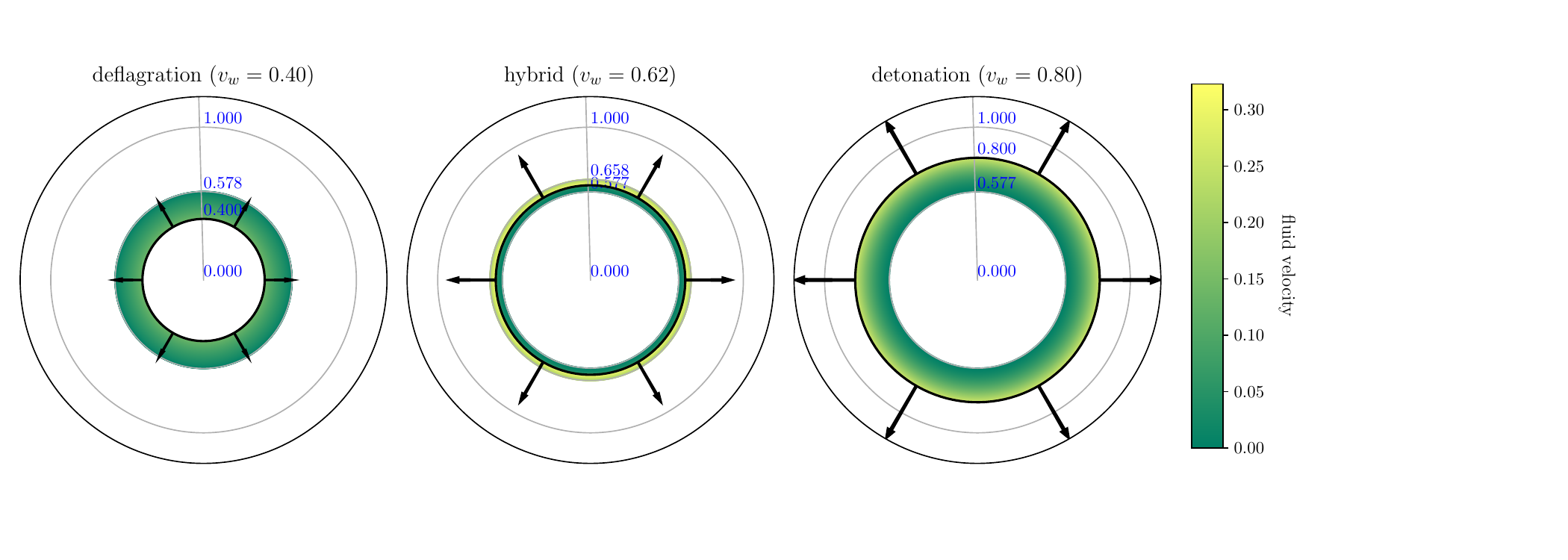}
	\caption{\small Depiction of different modes of combustion of a strong first order PT ($\alpha=0.1$). The colors show non-vanishing fluid flow $v_\mathrm{ip}(\xi)$ in the bubble center frame. The location of the wall $v_w$ is indicated by the black line. The shell thickness can be inferred from the radial coordinates $\xi_\mathrm{front}$ and $\xi_\mathrm{back}$ which are indicated by the blue numbers between 0 and~1. For a deflagration the bubble wall is preceeded by a shock, for a detonation the wall is followed by a rarefaction wave, hybrids are inbetween. For $\alpha\lesssim 0.3$ the fluid velocity $v_{\text{ip}}$ is non-relativistic, see e.g. fig 7 in ref. \cite{Espinosa:2010hh}.} 
	\label{fig:polarcombustion}
\end{figure}

%%%%%%%%%%%%%%%%%%%%%%%%%%%%%%%%%%%%%%%%%%%%%%%%%%%%%%%%%%
We follow~\cite{Hindmarsh:2019phv} to calculate the contribution to the GW energy density from colliding soundshells. We give a brief overview of the relevant equations for our analysis in the following. 
We are interested in the (dimensionful) power spectrum $\Omega^*_\mathrm{gw,ssm}(k)$ which is related to the dimensionless spectral density $P(k)$ via
\be
\Omega^*_\mathrm{gw,ssm}(k)=\frac{k^3}{2\pi^2}P_\mathrm{gw}(k)\, .
\ee
The computation of $\Omega^*_\mathrm{gw,ssm}(k)$ proceeds in three steps. First, the hydrodynamic equations of the plasma around a single bubble are solved, yielding the velocity profile $v_{\rm ip}(\xi)$ and enthalpy density $w_{\rm ip}(\xi)$ as a function of the coordinate $\xi = r/t$, with $r$ the radial displacement from the center of the bubble and $t$ the time since nucleation. One distinguishes between three different kinds of combustion, that are displayed in figure~\ref{fig:polarcombustion}. See~\cite{Espinosa:2010hh, Hindmarsh:2019phv} for the details of the solution. In fact, we are interested in the energy density contrast
\be
	\lambda = \frac{e - \bar e}{ \bar w}\,,
\ee
which is obtained from $w$ using the equation of state. The bars indicate that the quantities are evaluated at the nucleation temperature. From the invariant velocity profile and the density contrast we determine
\be
f(z)=\frac{4\pi}{z}\int_{0}^{\infty}\total\xi v_{\mathrm{ip}}(\xi)\sin(z\xi)\,\qquad l(z)=\frac{4\pi}{z}\int_{0}^{\infty}\total\xi \lambda_{\mathrm{ip}}(\xi)\xi\sin(z\xi)\,.
\ee
Both contribtutions can be combined into the spectral density for a single bubble
\be
|A(z)|^2=\frac{1}{4}\left[\left(f'(z)\right)^2+\left(c_s l(z)\right)^2\right].
\ee
This quantity enters in the second step, where the velocity spectrum is obtained by averaging over many bubbles. This requires a choice of the nucleation model. We choose the exponentially decreasing bubble lifetime distribution $\nu(\beta T)=\exp(-\beta T)$, one of the two choices of~\cite{Hindmarsh:2019phv}. Here, $T$ describes the time since the onset of the phase transition, and $\beta$ the nucleation rate. 

 The full velocity spectral density can then be expressed as integral over the single bubble spectrum, via
\be
\mathcal{P}_{\mathrm{\tilde{v}}}(q)=\frac{2}{(\beta R_*)^3}\frac{1}{2\pi}\left(\frac{q}{\beta}\right)^3\int\total\tilde{T}\nu(\tilde{T})\tilde{T}^6A|(\tilde{T}q/\beta)|^2
\label{eq:velocitypowerspectrum}
\ee
with $\tilde T\equiv \beta T$.

As a sidenote, the RMS fluid velocity follows from eq.~\ref{eq:velocitypowerspectrum}, which gives a contribution to the overall normalization factor of the GW energy density,\footnote{Note that this definition differs from Eq.~(\ref{eq:uf}). The difference between the two definitions is discussed in~\cite{Hindmarsh:2019phv} and turns out to be small numerically. }
\bea
\bar{U}_f^2&=&\int\total q\frac{1}{q}\mathcal{P}_{\mathrm{\tilde{v}}}(q)=\frac{2}{(\beta R_*)^3}\int\total\tilde{T}\nu(\tilde{T})\tilde{T}^3\int\total z\frac{z^2}{2\pi^2}|A(z)|^2 \nn \\
&=&\frac{3}{4\pi^2\xi_w^3}\int\total z\frac{z^2}{2\pi^2}2|A(z)|^2\, .
\eea
From this, note that the single bubble spectrum $|A(z)|^2$ contains essentially all necessary information to calculate the RMS fluid velocity.

For convenience it is useful to rescale the velocity power spectrum to
\be
P_v(q)=L_f^3\bar{U}_f^2\tilde{P}_v(qL_f),
\ee
where $L_f$ is the characteristic scale in the fluid.

\begin{figure}[t]
	\centering
	\includegraphics[width=0.99\linewidth]{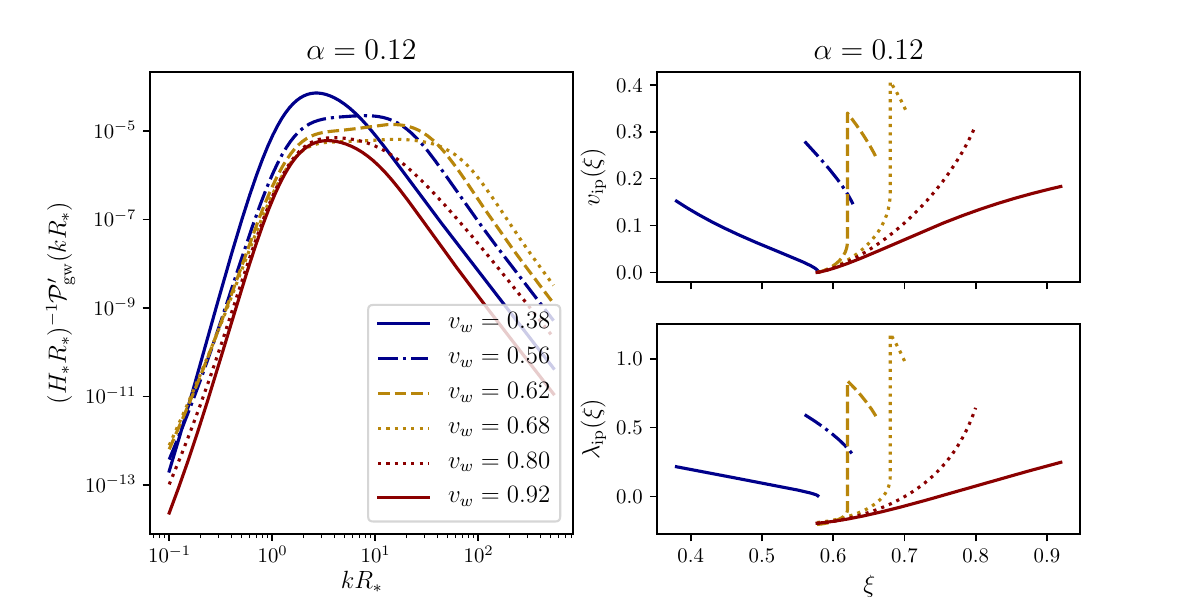}
	\caption{Left: Gravitational wave spectra as predicted by the soundshell model, before redshift. The PT strength is $\alpha=0.12$. For intermediate wall velocities, besides $R_*$, a second scale $L_{\rm ssh}$ emerges. Right: Corresponding invariant profiles, which enter the single-bubble spectrum $|A(z)|$.}
	\label{fig:soundshell01}
\end{figure}

Finally, the velocity spectrum can be correlated with itself to give the spectral density of GWs. The result for the relative growth rate of the GW power spectrum compared to the Hubble scale is
\be
\Omega^{'*}_{\mathrm{gw,ssm}}(k)=\frac{\dot{\Omega^*}_\mathrm{gw,ssm}}{H}=3\left(\Gamma\bar{U}_f^2\right)^2\left(H L_f\right)\frac{\left(kL_f\right)^3}{2\pi^2}\tilde{P}_{\mathrm{gw}}(k L_f)\, ,
\ee
with $\Gamma=\bar w/\bar e=4/3$ and
\be
\tilde{P}_{\mathrm{gw}}(q)=\frac{1}{4\pi y c_s}\left(\frac{1-c_s^2}{c_s^2}\right)^2\int_{z_-}^{z_+}\frac{\total z}{z}\frac{(z-z_+)^2(z-z_-)^2}{(z_++z_--z)}\tilde{P}_v(z)\tilde{P}_v(z_++z_--z)\, .
\ee 
Taking the lifetime estimate of the source into account, one has
\be
\Omega^*_{\mathrm{gw,ssm}}(k)=\left(H\tau\right)\Omega^{'*}_{\mathrm{gw,ssm}}(k)\, ,\qquad (H\tau)=\min(1,H\tau_{\mathrm{nl}})\, \label{eq:soundshellsource},
\ee
where $\tau_{\mathrm{nl}}=L_f/\bar{U}_f$ is the time scale where non-linearities evolve. Note that the rescaling freedom in $L_f$ removes the $\beta$ dependence from $\tilde{P}_{\rm gw}(y)$, such that the spectral information of the soundshell model is encoded by $\alpha$ and $v_w$ alone. The left panel of figure \ref{fig:soundshell01} shows soundshell spectra for fixed $\alpha=0.12$ and varying $v_w$. Multiple breaks can arise, depending on the type of combustion. One break position is universally set by $R_*$ which is the same for all profiles. For wall velocities close to the speed of sound, $c_s^2=1/3$, the second scale $L_{\rm ssh}$ emerges. 
For hybrids, close to the Jouguet velocity, the spectrum might even have an additional break, but we have not studied these systematically as these correspond to $\alpha<0.01$. Ultimately, these additional
features might be idiosyncratic to the sound shell model.

\newpage
\section{Fit results}\label{sec:appMCMC}
\begin{figure}[h]
	\centering
	\begin{overpic}[scale=.32]
	{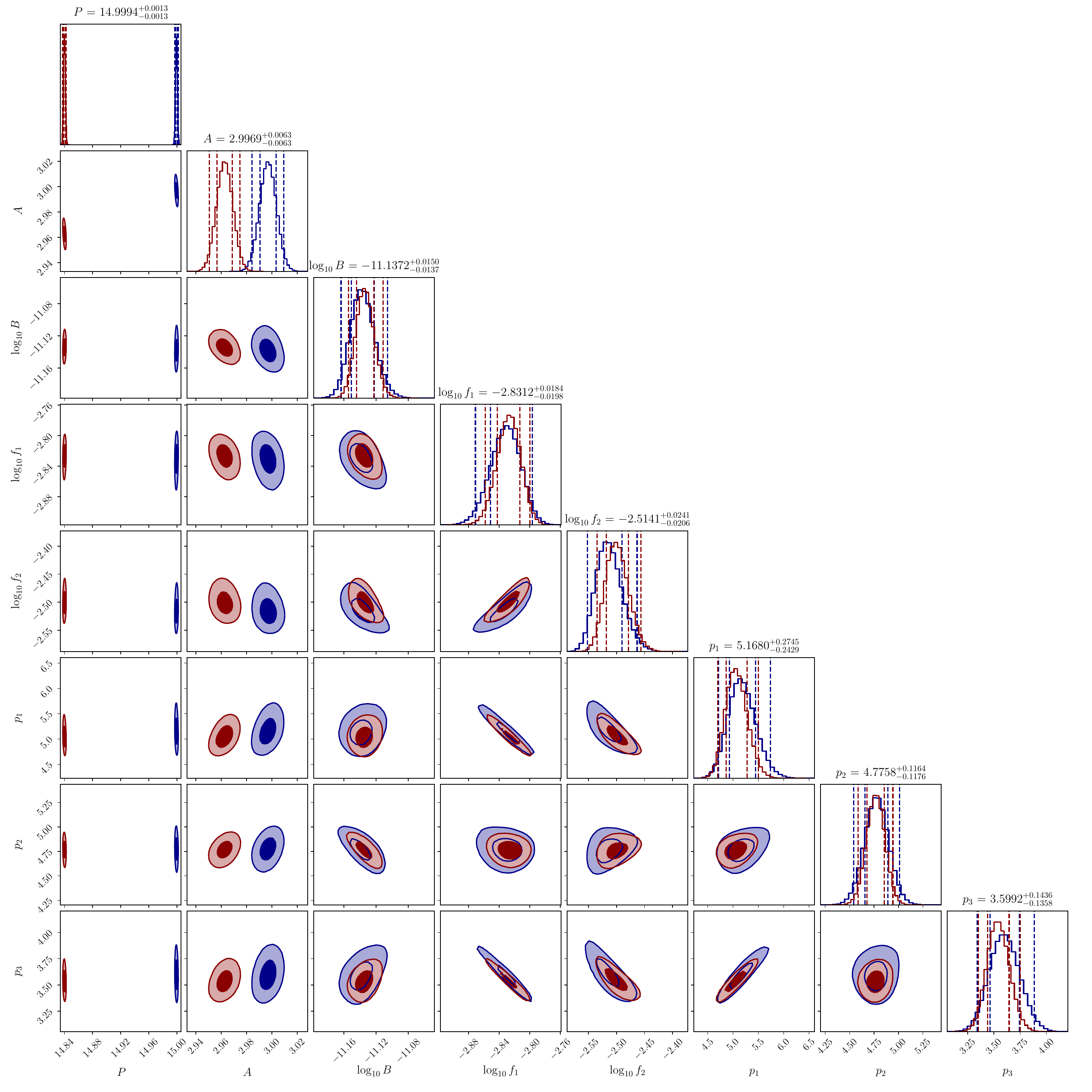}
	\put(51.5,61.5){\includegraphics[clip,trim=22 0 0 0,scale=0.5]{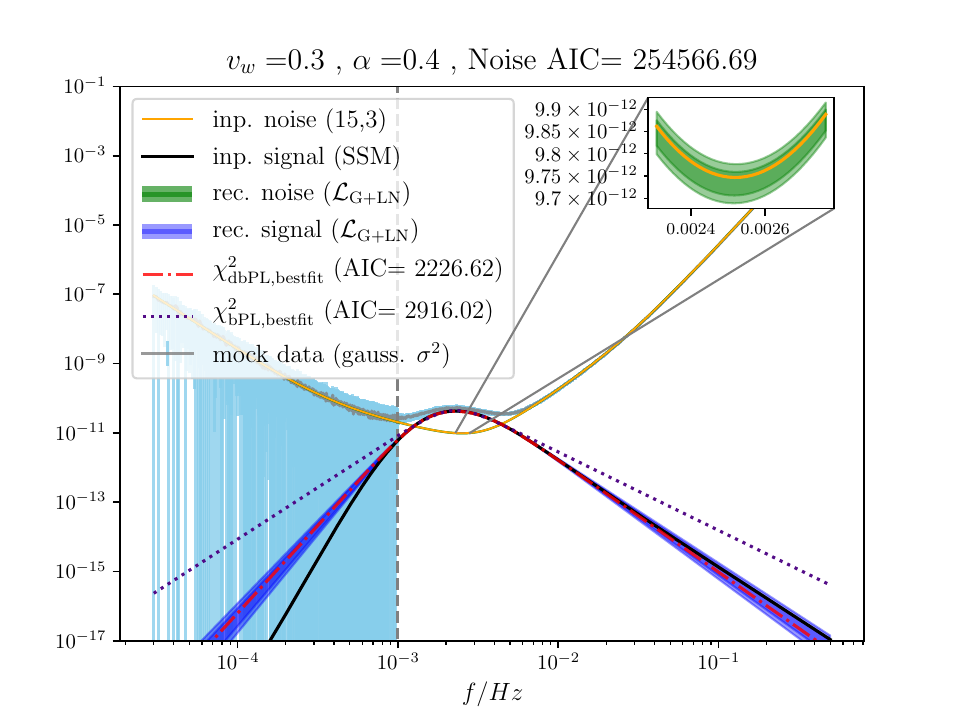}}
	\end{overpic}
	\caption{Double broken power law fit to the mock data, generated with the sound shell model for wall velocity $v_w=0.3$ and $\alpha=0.4$. Left: reconstruction of the double broken power law fit parameters. The red (blue) contours show the result corresponding to $\mathcal L_{\chi^2}$ ($\mathcal L_{\rm G+LN}$). Right: Corresponding mock data and noise and signal reconstruction as a function of frequency.
	} 
	\label{fig:triangle30alph40000}
\end{figure}

\begin{figure}[t]
	\centering
	\begin{overpic}[scale=.32]
		{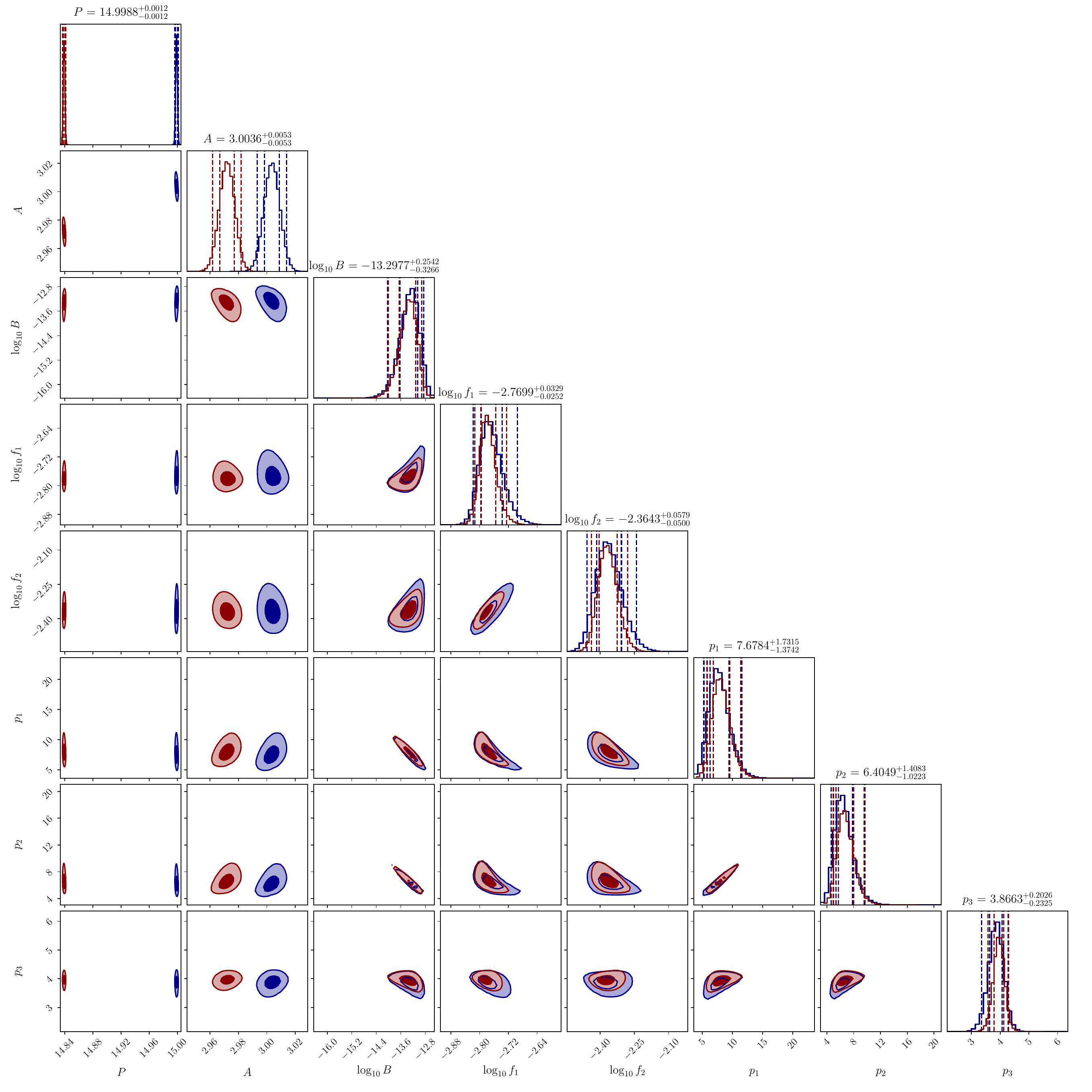}
		\put(51.5,61.5){\includegraphics[clip,trim=22 0 0 0,scale=0.5]{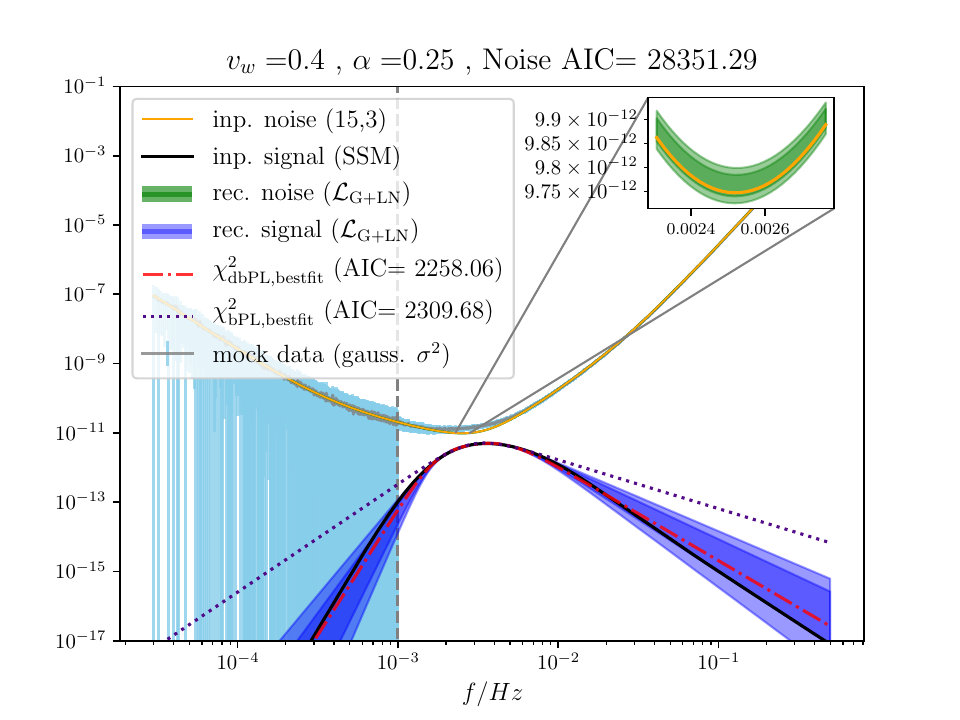}}
	\end{overpic}
	\caption{Double broken power law fit to the mock data, generated with the sound shell model for wall velocity $v_w=0.4$ and $\alpha=0.25$. Left: reconstruction of the double broken power law fit parameters. The red (blue) contours show the result corresponding to $\mathcal L_{\chi^2}$ ($\mathcal L_{\rm G+LN}$). Right: Corresponding mock data and noise and signal reconstruction as a function of frequency.}
	\label{fig:triangle40alph25000}
\end{figure}

\begin{figure}[t]
	\centering
	\begin{overpic}[scale=.32]
		{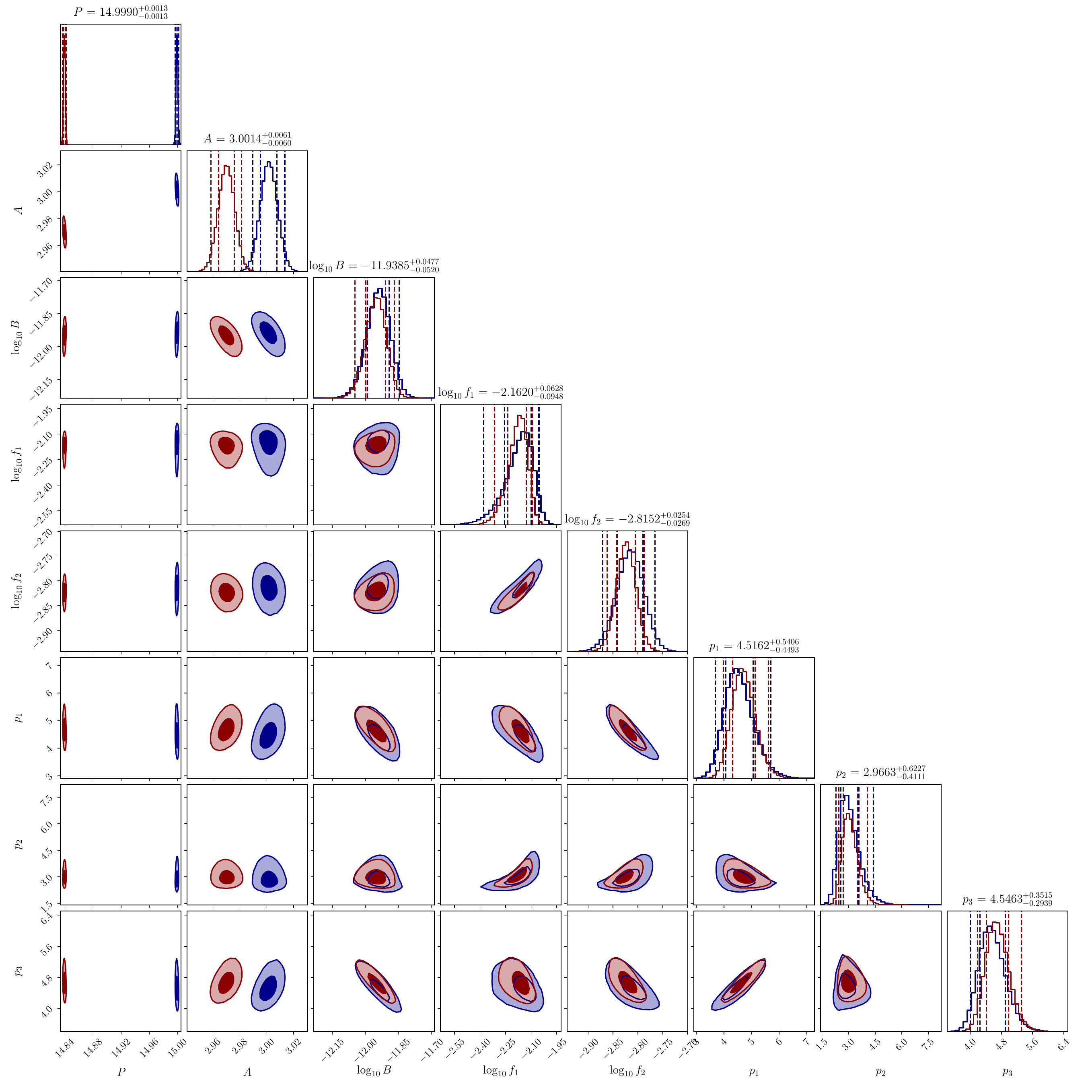}
		\put(51.5,61.5){\includegraphics[clip,trim=22 0 0 0,scale=0.5]{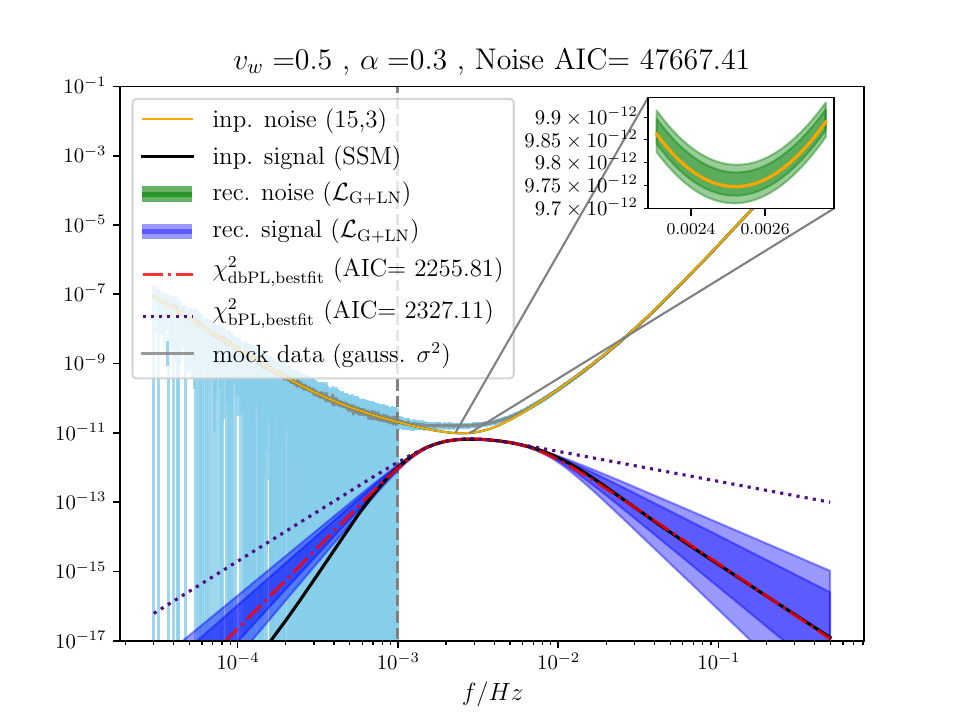}}
	\end{overpic}
	\caption{Double broken power law fit to the mock data, generated with the sound shell model for wall velocity $v_w=0.5$ and $\alpha=0.3$. Left: reconstruction of the double broken power law fit parameters. The red (blue) contours show the result corresponding to $\mathcal L_{\chi^2}$ ($\mathcal L_{\rm G+LN}$). Right: Corresponding mock data and noise and signal reconstruction as a function of frequency.}
	\label{fig:triangle50alph30000}
\end{figure}

\begin{figure}[t]
\centering
\begin{overpic}[scale=.32]
	{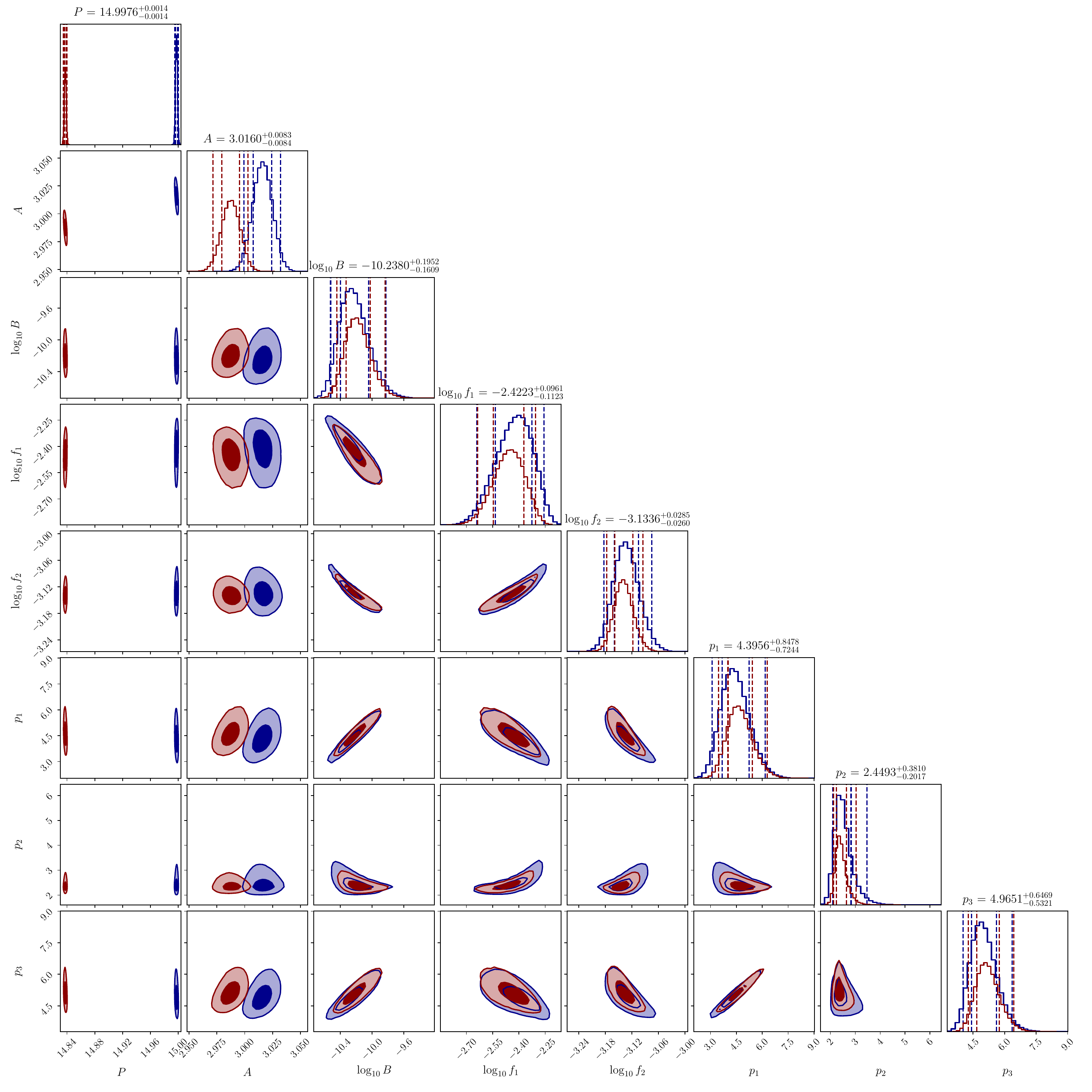}
	\put(51.5,61.5){\includegraphics[clip,trim=22 0 0 0,scale=0.5]{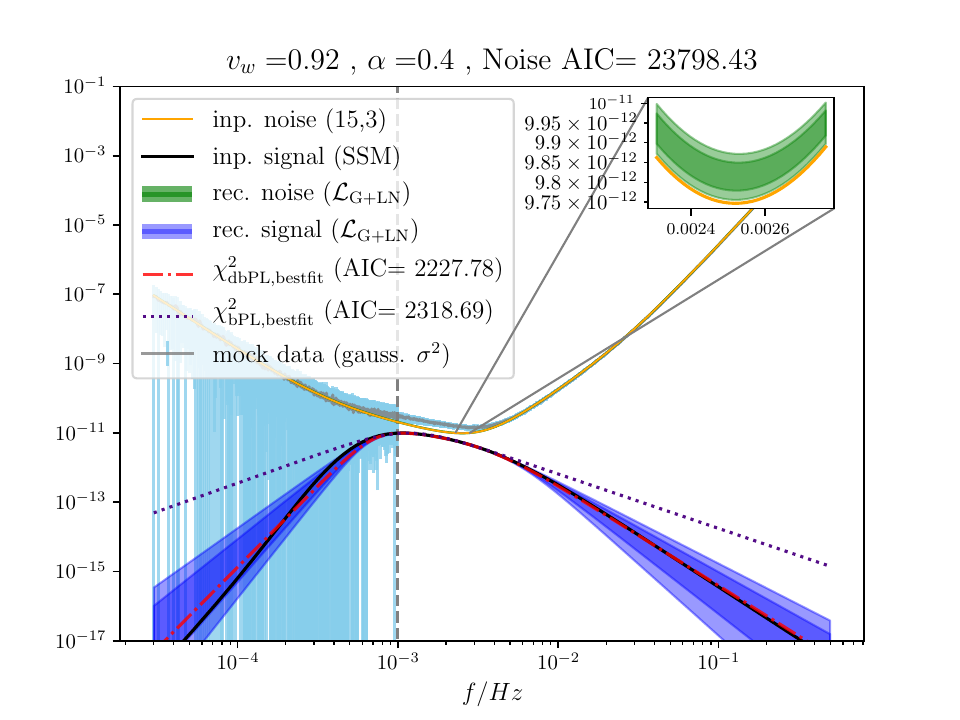}}
\end{overpic}
	\caption{Double broken power law fit to the mock data, generated with the sound shell model, for wall velocity $v_w=0.92$ and $\alpha=0.4$. Left: reconstruction of the double broken power law fit parameters. The red (blue) contours show the result corresponding to $\mathcal L_{\chi^2}$ ($\mathcal L_{\rm G+LN}$). Right: Corresponding mock data and noise and signal reconstruction as a function of frequency.}
\label{fig:triangle92alph40000}
\end{figure}

In this appendix, we show the details of the MCMC analysis of the data points discussed in Section~\ref{sec:MCMC}. Figures \ref{fig:triangle30alph40000} to \ref{fig:triangle92alph40000} show the results for the four benchmark points listed in Table~\ref{tab:MCMCbenchmarks}.

The upper right corner of each figure shows the (reconstructed) spectrum as a function of frequency. Here, we show the input noise signal with fixed $A=3$ and $P=15$, and the input gravitational wave signal, which is obtained from the soundshell model. The data with $f > 1 \, {\rm mHz}$ have been coarse-grained according to the procedure described in Section~\ref{sec:LISA}. In the frequency region $f<1 \, {\rm mHz}$, the data points are just given by the mean of 94 chunks. For some data points the spread in the data points is rather large, leading to error bands extending to the bottom of the graph. The finite resolution of the figure in combination with the large amount of data points makes this look rather dramatic, but in reality only 30\% of data points have such large error bars.  Note that we do not use the sample estimate of the standard deviation in the likelihood fit $\mathcal L _{\rm G+LN}$, but the uncertainty is estimated from the theoretical uncertainty of the fitted model following Eq.~(\ref{eq:cgfancy}). We show the reconstruction of noise (green) and signal (blue) with $\mathcal L _{\rm G+LN}$ with respective $1\sigma$ and $2\sigma$ credible intervals. For comparison, we also show the $\chi^2$ result of the (double) broken powerlaw fits in dashed red (dotted black), from which we estimate the fit quality via $\Delta\mathrm{AIC}$. The selected parameters all favor the double broken powerlaw fit by many units. The value of the AIC for pure noise fits is shown for reference on top of the panel and demonstrates that pure noise fits are heavily disfavored.

In the lower left panels we show the reconstruction of the double broken power law fit in parameter space\footnote{For easy visualization we use the package \cite{corner}.}. We compare the two choices for the likelihood functions from the main text, $\mathcal{L}_{\chi^2}$ (red) and $\mathcal{L}_\mathrm{G+LN}$ (blue). Concerning the noise (the two leftmost columns), one can see that switching to $\mathcal{L}_\mathrm{G+LN}$ removes the bias from the gaussian sample estimate of the uncertainties and the bias from the likelihood function. For the reconstruction of the signal (the six other columns), the statistical uncertainty dominates such that roughly the same central values are reconstructed using either likelihood. The $\chi^2$ likelihood tends to underestimate the uncertainties. 

\clearpage

%%%%%%%%%%%%%%%%%%%%%%%%%%%%%%%%%%%%%%%%%%%%%%%%%%%%%%%%%
%%%%%%%%%%%%%%%%%%%%%%%%%%%%%%%%%%%%%%%%%%%%%%%%%%%%%%%%%%
\newpage
\bibliographystyle{JHEP}
\bibliography{bib}

\end{document}